# False Discovery Rate Control for High-Dimensional Networks of Quantile Associations Conditioning on Covariates

Jichun Xie[a,*], Ruosha Li[b]

[a] *Department of Biostatistics and Bioinformatics, Duke University School of Medicine, Durham, NC 27707*

[b] *Department of Biostatistics and Data Science, the University of Texas Health Science Center at Houston, Houston, TX 77030*

---

**Abstract**

Motivated by the gene co-expression pattern analysis, we propose a novel Sample QUAntile-based Contingency (SQUAC) statistic to infer quantile associations conditioning on covariates. It features enhanced flexibility in handling variables with both arbitrary distributions and complex association patterns conditioning on covariates. We first derive its asymptotic null distribution, and then develop a multiple testing procedure based on SQUAC to simultaneously test the independence between one pair of variables conditioning on covariates for all $p(p-1)/2$ pairs. Here, $p$ is the length of the outcomes and could exceed the sample size. The testing procedure does not require resampling or perturbation, and thus is computationally efficient. We prove by theory and numerical experiments that this testing method asymptotically controls the false discovery rate (FDR). It outperforms all alternative methods when the complex association panterns exist. Applied to a gastric cancer data, this testing method successfully inferred the gene co-expression networks of early and late stage patients. It identified more changes in the networks which are associated with cancer survivals. We extend our method to the case that both the length of the outcomes and the length of covariates exceed the sample size, and show that the asymptotic theory still holds.

---

*Principal corresponding author

*Email addresses:* `jichun.xie@duke.edu` (Jichun Xie), `ruosha.li@uth.tmc.edu` (Ruosha Li)

## 1. Introduction

In recent years, inference on high-dimensional association networks has attracted considerable attention. Let $\boldsymbol{Y}$ represent a high-dimensional multivariate outcome. The goal is to model the association between elements of $\boldsymbol{Y}$, without or with covariates $\boldsymbol{X}$. Linear associations or rank associations have been studied in many existing literature. See, for example, Drton and Perlman (2007), Li et al. (2012), Cai et al. (2013b), Chen et al. (2016), and Cai and Liu (2016). Beyond linear or rank associations, more general association without parametric distribution assumptions have been studied. Chun et al. (2016) used joint quantile regression with penalization to estimate graphical models. Voorman et al. (2014) proposed an additive model to describe the relationship between variables, and estimated their graphical structure using basis function expansion and penalized regression. Li et al. (2014a) proposed a similar model to study the graphical structure induced by additive conditional independence, and focused on its theoretical features. Although these methods modeled general associations, they do not allow for adjusting for covariates $\boldsymbol{X}$. In addition, their general association estimators have relative complex forms so that their asymptotic null distributions are hard to derive. As a result, it is hard to control the type I error when inferring general associations.

An example of the general association network is the gene co-expression network. In gene co-expression network analysis, high-dimensional gene expression data (microarray or RNA-seq) are collected, often times together with other covariates. Three main challenges, among others, exist in the analysis. First, covariates (such as gender, race, cancer stage, *etc.*) may affect the expression distributions or distort their association patterns. Therefore, ignoring these covariates will lead to both false discoveries and false non-discoveries in network inference. Second, gene expressions cannot always be normalized into Gaussian or commonly seen parametric distributions. To properly model the co-expression pattern, we need a model that allows flexible expression distributions and association patterns. Third, a type I error measure is often desired for the analysis so that researchers can understand how reliable the inferred associations are and prioritize the validation studies. To the best of our knowledge, no existing method can address these three challenges simultaneously.

To address these challenges, we develop a robust and computationally efficient multiple testing procedure to infer a high-dimensional sparse general association network conditioning on covariates. Our method uses conditional quantile associations to capture a wide range of general association patterns. The proposed test statistic (SQUAC) has a neat asymptotic null distribution, allowing us to conduct computationally efficient multiple testing with accurate FDR control.

The rest of the paper is organized as follows. In Section 2, we propose a novel summary statistic to evaluate the pairwise quantile association conditioning on covariates. We also propose a multiple testing procedure that controls the FDR. In Section 3, we prove the test statistic converges to the chi-square distribution under the null, and the FDR control procedure is valid. In Section 4, we design numerical experiments to compare the proposed multiple testing method and several alternative methods. A real data analysis is conducted in Section 5 to investigate the dependence patterns among patients with early stage and late stage gastric cancer. In Section 6, we discuss the case with high-dimensional sparse covariates. Further discussions are provided in Section 7.

## 2. Method

The following notations are used in the paper. For any $a \in \mathbb{R}$, $\lfloor a \rfloor$ represents the largest integer that is smaller than or equal to $a$. For two numbers $a, b \in \mathbb{R}$, $a \vee b = \max(a, b)$, and $a \wedge b = \min(a, b)$. For a vector $\mathbf{a} = (a_1, \ldots, a_n)' \in \mathbb{R}^n$, let $\|\mathbf{a}\|_p = (\sum_{i=1}^n a_i^p)^{1/p}$, for any $p > 0$. Also let $\|\mathbf{a}\|_0 = \sum_{i=1}^n I(a_i \neq 0)$. For a symmetric matrix $A$, denote by $\lambda_{\max}(A)$ its largest eigenvalue. For a set $\mathbb{A}$, let $\mathsf{Card}(\mathbb{A})$ represent its cardinality. For two sequences of real numbers $\{a_n\}$ and $\{b_n\}$, write $a_n = O(b_n)$ if there exists a constant $C$ such that $a_n \leq C b_n$ holds for all sufficiently large $n$, write $a_n = o(b_n)$ if $\lim_{n\to\infty} a_n/b_n = 0$. If $a_n = O(b_n)$ and $b_n = O(a_n)$, then $a_n \asymp b_n$. If $\lim_{n\to\infty} a_n/b_n = 1$, write $a_n \sim b_n$. For a sequence of random variables $\{X_n\}$, and a sequence of real numbers $\{a_n\}$, if there exists a constant $M$ such that for any $\varepsilon > 0$ $\lim_{n\to\infty} \mathsf{P}(|X_n/a_n| > M) < \varepsilon$ for sufficiently large $n$, write $X_n = O_p(a_n)$.

### *2.1. Sample Quantile-Based Contingency Table Statistics*

Let $\boldsymbol{Y} = (Y_1, \ldots, Y_p)'$ be the outcome vector and $\boldsymbol{X} = (X_1, \ldots, X_{p_x})'$ be the covariate vector. For convenience, assume $X_1 \equiv 1$ is the intercept term. We are interested to test the null hypotheses: $\mathrm{H}_{0,ij} : Y_i \perp\!\!\!\perp Y_j \mid \boldsymbol{X}$ *vs.* $\mathrm{H}_{1,ij} : Y_i \not\perp\!\!\!\perp Y_j \mid \boldsymbol{X}$. In a special case where no covariates exists, let $X = 1$ and then the null hypothesis becomes $\mathrm{H}_{0,ij} : Y_i \perp\!\!\!\perp Y_j$.

Motivated by the commonly used Pearson's chi-square independence test, we propose a new quantile association test based on quantile regressions. Let $0 = \tau_0 < \tau_1 < \ldots < \tau_{D-1} < \tau_D = 1$ be an increasing sequence of probability points. The probability jump between the two quantile points is $\nu_s = \tau_s - \tau_{s-1}$, for $s \in \{1, \ldots, D\}$. Let $Q_{i,s}$ be the $\tau_s$-th conditional quantile of $Y_i$ given $\boldsymbol{X}$. For convenience, we set $Q_{i,0} = -\infty$ and $Q_{i,D} = \infty$. We model the relationship between $Y_i$ and $\boldsymbol{X}$ by a quantile regression model,

$$Q_{i,s} = \boldsymbol{X}'\boldsymbol{\beta}_{0,i}(\tau_s), \quad s \in \{1, \ldots, D-1\}. \tag{1}$$

Quantile regression is more flexible than linear regression, because the coefficients $\boldsymbol{\beta}_{0,i}(\tau_s)$ can differ across different quantile levels $\tau_s$. Suppose for the $k$th subject, we observe $\boldsymbol{Y}_k = (Y_{k1}, \ldots, Y_{kp})'$ and $\boldsymbol{X}_k = (X_{k1}, \ldots, X_{kp_x})'$ as a realization of the outcome and the covariate, $k \in \{1, \ldots, n\}$. When $p_x << n$, the quantile regression coefficients can be estimated consistently (Koenker, 2005) by

$$\hat{\boldsymbol{\beta}}_i(\tau_s) = \arg\min_{\beta \in \mathbb{R}^{p_x}} \sum_{k=1}^{n} \{\rho_{\tau_s}(Y_{ki} - \boldsymbol{X}'_k\boldsymbol{\beta})\}, \quad s \in \{1, \ldots, D-1\} \tag{2}$$

with the loss function $\rho_{\tau_s}(y) = y\{\tau_s - I(y < 0)\}$. Subsequently, the conditional $\tau_s$-th sample quantile is estimated by $\hat{Q}^*_{k,i,s} = \boldsymbol{X}'_k\hat{\boldsymbol{\beta}}_i(\tau_s)$, for $s \in \{1, \ldots, D-1\}$, $\hat{Q}^*_{k,i,0} = -\infty$, and $\hat{Q}^*_{k,i,D} = \infty$.

In reality, the randomness in quantile fitting might result in quantile level crossing problems: there might exist some $\hat{Q}^*_{k,i,s} > \hat{Q}^*_{k,i,s+1}$, for some $k \in \{1, \ldots, n\}$, $i \in \{1, \ldots, p\}$, $s \in \{1, \ldots, D-2\}$. Many papers discussed how to modify the quantile regression fitting to obtain non-crossing estimated quantiles. See, for example, He (1997), Neocleous and Portnoy (2008), and Bondell et al. (2010). In this paper, to minimize the conditions needed and simplify the computation procedure, we propose a swapping approach to solve the possible quantile level crossing problem.

After deriving the estimated conditional quantiles $\hat{Q}^*_{k,i,s} = \boldsymbol{X}'_k\hat{\boldsymbol{\beta}}_i(\tau_s)$, for $s \in \{1, \dots, D-1\}$, we start from the beginning of the sequence and check the adjacent estimators. For $s = 1, \dots, D-2$, if $\hat{Q}^*_{k,i,s} > \hat{Q}^*_{k,i,s+1}$, we swap them so that $\hat{Q}_{k,i,s} = \hat{Q}^*_{k,i,s+1}$ and $\hat{Q}_{k,i,s+1} = \hat{Q}^*_{k,i,s}$; otherwise let $\hat{Q}_{k,i,s} = \hat{Q}^*_{k,i,s}$ and $\hat{Q}_{k,i,s+1} = \hat{Q}^*_{k,i,s+1}$. In this way, the estimated quantile levels will not cross. In Section 3, we will show that the swapping approach will maintain the theoretical properties needed for proving the theoretical results in the paper.

For each pair of the outcome variables $(Y_i, Y_j)$, we then construct a $D \times D$ Sample QUAntile based Contingency (SQUAC) table. The $(s,t)$th cell of the table is the counts

$$O_{ij,st} = \sum_{k=1}^{n} I\{\hat{Q}_{k,i,s-1} < Y_{ki} \le \hat{Q}_{k,i,s}\} I\{\hat{Q}_{k,j,t-1} < Y_{kj} \le \hat{Q}_{k,j,t}\}.$$

It is easy to see that if all the conditional quantiles $Q_{k,i,s}$ are known, under $\mathrm{H}_{0,ij}$, the expected number of observations that will fall in the $(s,t)$th cell is $E_{st} = n\nu_s\nu_t$. Inspired by the Pearson chi-square test, we proposed the SQUAC test statistic

$$T_{ij} = \sum_{s=1}^{D}\sum_{t=1}^{D} \frac{(O_{ij,st} - E_{st})^2}{E_{st}}. \tag{3}$$

Although $T_{ij}$ and the Pearson chi-square test statistic look similar in format, they are fundamentally different. The traditional Pearson chi-square test statistic is designed for testing independence between categorical variables, so that the cell boundaries of the contingency table are pre-fixed. For a SQUAC table, because the cell boundaries depend on quantile regression estimators, they are different across samples, and for each sample, the cell boundaries vary when others samples change. When deriving the asymptotic properties of $T_{ij}$ and the testing procedure, we will include the variation introduced by the conditional quantile point estimation.

### *2.2. False Discovery Rate Control Procedure*

By bounding the errors between the estimated conditional quantiles and the true quantiles and projection techniques, we will show that under $\mathrm{H}_{0,ij}$, $T_{ij}$ asymptotically follows $\chi^2\{(D-1)^2\}$. Under the $\mathrm{H}_{1,ij}$, for any constant $C$, as $n$ and $p$ increase, the probability that $T_{ij} \leq C$ will be vanishing (See Theorem 1, Corollary 2, and their proofs for details). Based on these results, we construct a false discovery rate (FDR) control procedure to test $\mathrm{H}_{0,ij}$ $(1 \leq i < j \leq p)$ simultaneously. We reject $\mathrm{H}_{0,ij}$ if $T_{ij} > t$, and determine the threshold $t$ based on the desired FDR level $\alpha$.

Define the null set $\mathbb{H}_0 = \{(i,j) : 1 \leq i < j \leq p,\ \mathrm{H}_{0,ij} \text{ is true}\}$, and the full set $\mathbb{H} = \{(i,j) : 1 \leq i < j \leq p\}$. The alternative set is $\mathbb{H}_1 = \mathbb{H} \setminus \mathbb{H}_0$. Let

$$q = \mathsf{Card}(\mathbb{H}) = p(p-1)/2 \quad \text{and} \quad q_0 = \mathsf{Card}(\mathbb{H}_0). \tag{4}$$

The false discovery proportion (FDP) and FDR are defined as

$$\text{FDP} = \frac{\sum_{(i,j)\in\mathbb{H}_0} I(T_{ij} > t)}{\max\{\sum_{(i,j)\in\mathbb{H}} I(T_{ij} > t), 1\}} = \frac{\sum_{(i,j)\in\mathbb{H}_0} I(T_{ij} > t)/q_0}{\max\{\sum_{(i,j)\in\mathbb{H}} I(T_{ij} > t), 1\}/q} \cdot \frac{q_0}{q}, \quad \text{FDR} = \mathsf{E}(\text{FDP}). \tag{5}$$

Since $\mathbb{H}_0$ is not known, we replace $\sum_{(i,j)\in\mathbb{H}_0} I(T_{ij} > t)/q_0$ with $G_D(t)$, the complementary CDF of $\chi^2\{(D-1)^2\}$. We further assume $q_0/q \to 1$ as $p \to \infty$. It then leads to the following testing procedure.

Let $t_p = 4\log(n \vee p) + \{(D-1)^2 - 2\}\log\log(n \vee p)$, and

$$\hat{t} = \inf\left\{0 \leq t \leq t_p : \frac{qG_D(t)}{\max\left\{\sum_{1\leq i<j\leq p} I\{T_{ij} > t\}, 1\right\}} \leq \alpha.\right\}. \tag{6}$$

If $\hat{t}$ does not exist, let $\hat{t} = t_p$. For $1 \leq i < j \leq p$, reject $\mathrm{H}_{0,ij}$ if $T_{ij} > \hat{t}$.

The searching range of $\hat{t}$ is restricted to $[0, t_p]$ because by the proof of Theorem 3, $\lim_{n,p\to\infty} \mathsf{P}(\sup_{(i,j)\in\mathbb{H}_0} T_{ij} > t_p) = 0$. In practice, to find $\hat{t}$, we only need to search at the realization values of that $T_{ij}$ in the interval $[0, t_p]$.

We provided an equivalent form in Algorithm 1. In Algorithm 1, $\widehat{\text{FDR}}_\ell$ can be viewed as a scaled $p$-value. If $\widehat{\text{FDR}}_\ell \leq \alpha$, the surrogate $p$-value $G_D(T_{h(\ell)}) \leq (r+\ell)\alpha/q$. Therefore, Algorithm 1 is essentially a Benjamini-Hochberge type procedure (Benjamini and Hochberg, 1995) with pre-thresholding. The computation complexity is $O\{Dnp + p^2\log(n \vee p)\}$. Considering that the total number of pairs $(Y_i, Y_j)$ is $p(p-1)/2 = O(p^2)$, this algorithm is very efficient. To facilitate the general audience to use this method, we provided the R codes, the corresponding help file, and some examples at https://github.com/jichunxie/squac.

**Algorithm 1:** FALSE DISCOVERY RATE CONTROL ALGORITHM.

**Input**: SQUAC statistics $T_{ij}$, $(i,j) \in \mathbb{H}$.

Set the rejection set $\mathcal{R} = \emptyset$, the candidate set $\mathcal{V} = \emptyset$, and the exist label state $= 0$.

**For** $(i,j) \in \mathbb{H}$**:**

**If** $T_{ij} > t_p$**:** $\mathcal{R} = \mathcal{R} \cup \{(i,j)\}$;

**else:** $\mathcal{V} = \mathcal{V} \cup \{(i,j)\}$.

Set $r = \mathsf{Card}(\mathcal{R})$.

Rank $T_{h(1)} \geq T_{h(2)} \geq \ldots \geq T_{h(q-r)}$, where $h : \{1, \ldots, q-r\} \to \mathcal{V}$ is the corresponding index mapping.

Set $\ell^* = 0$.

**While** $\ell \in \{q-r, \ldots, 1\}$ *and state* $= 0$**:**

Set $\text{FDR}_\ell = qG_D(T_{h(\ell)})/(r+\ell)$.

**If** $\text{FDR}_\ell \leq \alpha$**:** Set $\ell^* = \ell$, and state $= 1$.

Set $\mathcal{R} = \mathcal{R} \cup \{h(1), \ldots, h(\ell^*)\}$.

**Output**: The rejection set $\mathcal{R}$.

## 3. Asymptotic Properties

For any $1 \leq i \leq p$, define $s_0(i) = \mathsf{Card}\{j : 1 \leq j \leq p, j \neq i, Y_i \not\perp\!\!\!\perp Y_j \mid \boldsymbol{X}\}$, and $d_p = \max_{1 \leq i \leq p} s_0(i)$. If we treat each variable $Y_i$ as a node and the dependence between a pair of variables as an edge, in the resulting network, $d_p$ is the maximum degree. To capture the association level in the $(s,t)$th cell of the SQUAC table, let

$$\gamma_{ij,st} = \mathsf{E}[\{I(Q_{k,i,s-1} < Y_{ki} \leq Q_{k,i,s}) - \nu_s\}\{I(Q_{k,j,t-1} < Y_{kj} \leq Q_{k,j,t}) - \nu_t\}/\{\nu_s\nu_t\}^{1/2}].$$

It is easy to see that $\gamma_{ij,st} = 0$ if $Y_i$ and $Y_j$ are independent given $\boldsymbol{X}$. For any constant $M > 0$, define the high quantile association set as

$$\mathbb{B}_M = \{(i,j) : \; 1 \leq i,j \leq p, \; \exists (s,t), \text{ such that } |\gamma_{ij,st}| \geq M\{\log(n \vee p)/n\}^{1/2}\}. \qquad (7)$$

To derive the asymptotic properties, we need the following conditions:

C1. Let $f_{\boldsymbol{X},i}$ be the conditional probability density function (PDF) of $Y_i \mid \boldsymbol{X}$. Assume $|f_{\boldsymbol{X},i}(y)| \le C_1$, $\forall\, y \in \mathbb{R}$, $\boldsymbol{X} \in \mathbb{R}^{p_x}$, and $i = 1, \ldots, p$.

C2. Assume $p_x$ is a constant. And also there exists a constant $C_2$ and some small positive constant $\varepsilon$ such that

$$\mathsf{P}\left[\sup_{s=1}^{D}|\hat{Q}_{k,i,s} - Q_{k,i,s}| > 2C_2(1+\varepsilon)n^{-1/2}\{\log(n \vee p)\}^{1/2}\right] < D(n \vee p)^{-2-\varepsilon}. \tag{8}$$

C3. Suppose $p \le cn^r$, for some $r > 0$. Also suppose there exists some $C_3 > 0$ such that $D \le C_3 \left\{\frac{\log(n \vee p)}{\log\log(n \vee p)}\right\}^{1/2}$.

C4. Let $0 < u_0 = \min_{1\le s\le D} \nu_s \le \max_{1\le s\le D} \nu_s = u_1 < 1$. Define constants $M_1 = (1 - u_0)(1-u_0+u_1)/u_0$, $M_2 = (3-2u_0)(4C_1C_2+2)/u_0$, and $M_3 = 4.02+2M_1+M_2$. Consider $\mathbb{B}_M$ defined in (7) with $M = M_3$. Let $c_p = \mathsf{Card}(\mathbb{B}_{M_3})$. We assume $d_p = o(\sqrt{c_p})$.

Condition C1 requires the PDF of $Y_i \mid \boldsymbol{X}$ to be bounded. It is a mild condition satisfied by a large family of distributions.

Condition C2 requires the estimated quantile levels achieve certain rate of convergence. This condition is also easily satisfied. In traditional quantile regression, under mild conditions, the Bahadur representation $n^{1/2}\{\hat{\boldsymbol{\beta}}_i(\tau) - \boldsymbol{\beta}_{0,i}(\tau)\} = \mathbf{D}_i(\mathbf{X},\tau)\mathbf{W}_i(\tau) + \mathbf{R}_i$ holds, where $\mathbf{D}_i(\mathbf{X},\tau) = \Big\{E\Big(\mathbf{X}\mathbf{X}' f_{\boldsymbol{X},i}\{\mathbf{X}'\boldsymbol{\beta}_{0,i}(\tau)\}\Big)\Big\}^{-1}$, $\mathbf{W}_i(\tau)$ is a Brownian Bridge, and $\mathbf{R}_i$ is an error term. It satisfies that for any constant $K > 0$,

$$\mathsf{P}\Big\{\|\mathbf{R}_i\|_2 = O(n^{-1/2}(\log n)^{3/2})\Big\} \ge n^{-K}. \tag{9}$$

See Result 1 in the Appendix of Portnoy (2012).

If $\sup_{k=1}^{n}\|\boldsymbol{X}_k\|_2 \le C$, $\sup_{i=1}^{p}\sup_{s=1}^{D}\lambda_{\max}(\mathbf{D}_i(\boldsymbol{X},\tau_s)) \le C$, and $p \le cn^r$ for some $r > 0$,

then for some sufficiently large $C_2$ we have

$$\begin{aligned}
&\mathsf{P}\left[|\hat{Q}^*_{k,i,s} - Q_{k,i,s}| > 2C_2(1+\varepsilon)n^{-1/2}\{\log(n \vee p)\}^{1/2}\right] \\
&= \mathsf{P}\left[n^{1/2}|\boldsymbol{X}'_k\hat{\boldsymbol{\beta}}_i(\tau_s) - \boldsymbol{X}'_k\boldsymbol{\beta}_{i,0}(\tau_s)| > 2C_2(1+\varepsilon)\{\log(n \vee p)\}^{1/2}\right] \\
&\leq \mathsf{P}\left[\|\boldsymbol{X}_k\|_2\|\mathbf{D}_i(\mathbf{X},\tau_s)\mathbf{W}_i(\tau_s) + \mathbf{R}_i\|_2 > 2C_2\{(1+\varepsilon)\log(n \vee p)\}^{1/2}\right] \\
&\leq \mathsf{P}\left[\frac{C\|\mathbf{D}_i(\mathbf{X},\tau_s)\mathbf{W}_i(\tau_s)\|_2}{C_2} > 2\{(1+\varepsilon)\log(n \vee p)\}^{1/2} - 1\right] + \mathsf{P}\left\{\|\mathbf{R}_i\|_2 > \frac{C_2}{C}\right\} \qquad (10) \\
&< 2(n \vee p)^{-2-2\varepsilon} + (n \vee p)^{-2-2\varepsilon} < (n \vee p)^{-2-\varepsilon} \qquad (11)
\end{aligned}$$

The term (10) leads to (11) because when $C_2$ is sufficiently large, $\|\mathbf{D}_i(\mathbf{X},\tau_s)\mathbf{W}_i(\tau_s)\|_2/C_2$ follows a Gaussian distribution with mean 0 and variance less than or equal to 1. Then by Gaussian tail probability, we have the large deviation probability controlled for the first part. For the second part, we use (9) and take $K = (2+2\varepsilon)(r \vee 1)$. It then follows

$$\mathsf{P}\left[\sup_{s=1}^{D}|\hat{Q}^*_{k,i,s} - Q_{k,i,s}| > 2C_2(1+\varepsilon)n^{-1/2}\{\log(n \vee p)\}^{1/2}\right] < D(n \vee p)^{-2-\varepsilon}, \qquad (12)$$

where we define $\hat{Q}^*_{k,i,D} = Q_{k,i,D} = \infty$ and $\hat{Q}^*_{k,i,D} - Q_{k,i,D} = 0$. Although we considered the convergence of $\hat{Q}^*_{k,i,s}$ under the conditions

$$\sup_{k=1}^{n}\|\boldsymbol{X}_k\|_2 \leq C \quad \text{and} \quad \sup_{i=1}^{p}\sup_{s=1}^{D}\lambda_{\max}(\mathbf{D}_i(\mathbf{X},\tau_s)) \leq C,$$

these conditions are sufficient instead of necessary. It is possible that under weaker conditions, (12) still holds. For example, the condition $\sup_{k=1}^{n}\|\boldsymbol{X}_k\|_2 \leq C$ can be weakened as $\mathsf{P}\left(\sup_{k=1}^{n}\|\boldsymbol{X}_k\|_2 \leq C\right) > 1 - n^{-3r}$. In Section 2, we proposed a swapping approach to guarantee that the estimated quantiles will be non-crossing. After swapping, $|\hat{Q}_{k,i,s} - Q_{k,i,s}| \leq \sup_{t\in\{1,\ldots,D-1\}}|\hat{Q}^*_{k,i,t} - Q_{k,i,t}|$. Thus, (12) also holds for $\hat{Q}_{k,i,s}$.

Condition C2 assumes $p_x$ is finite as $n$ goes to infinity. We will discuss $p_x \geq n$ case in Section 6.

Condition C3 allows $p$ to grow faster than $n$, so that the asymptotic properties hold for high-dimensional data. It also indicates that when $n$ is sufficiently large, $D$ can also be relatively large so that the SQUAC statistics can capture subtle conditional quantile associations in a local region of $(Y_i, Y_j)$.

Condition C4 requires the preset quantile levels to be bounded away from 0 and 1, because when quantile levels approaches to 0 or 1, the convergence rate of the estimated conditional quantiles becomes slower. Although it can be relaxed to allow the quantile levels approaching to 0 or 1, we keep it this way to make the presentation of the asymptotical theory simple. Condition C4 also requires a subset of dependent variables to have high quantile associations. It does not impose any lower bound condition on the minimum nonzero quantile association. The number 4.02 in the constant $M_3$ can be replaced by any constant greater than 4. The condition $d_p = o(\sqrt{c_p}) \leq o(p)$ leads to $q_0/q \to 1$ when $p \to \infty$. Here $q_0$ and $q$ are defined in (4). The term $q_0/q$ represents the sparsity level of the conditional quantile association network of $(Y_1, \ldots, Y_p) \mid \boldsymbol{X}$. When $c_p = O(p^\alpha)$ for some $\alpha \in (1, 2)$, $d_p$ could be much larger than $p^{1/2}$.

**Theorem 1.** *For any positive integer $D$, let $G_D(t)$ be the complementary* CDF *of $\chi^2\{(D-1)^2\}$. Under Conditions C1-C3, for any constant $C_0 > 0$,*

$$\sup_{0 \leq t \leq C_0 \log(n \vee p)} \sup_{(i,j) \in \mathbb{H}_0} \frac{P(T_{ij} \geq t)}{G_D(t)} \to 1. \tag{13}$$

**Corollary 1.** *Suppose Conditions C1 and C2 hold. For any positive integer $D$, under the null $\mathrm{H}_{0,ij}$ : $Y_i \perp\!\!\!\perp Y_j \mid \boldsymbol{X}$, the* SQUAC *statistic $T_{ij}$ asymptotically follows the chi-square distribution with the degree of freedom $(D-1)^2$.*

Theorem 1 considers the fact that the conditional quantile points are estimated rather than known. It shows that the true complementary CDF function of $T_{ij}$ converges to the chi-square complementary CDF over the range $[0, C_0 \log(n \vee p)]$ for all $(i, j) \in \mathbb{H}_0$. The convergence at the tail is a stronger result than the general convergence in distribution. At the tail point $C_0 \log(n \vee p)$, the complementary CDF of the chi-square distribution is very small. The convergence rate has to be faster than the decaying rate of the complementary CDF itself to make it happen.

To prove Theorem 1, we decompose $T_{ij,st} = \tilde{T}_{ij,st} + R_{ij,st}$, where

$$\tilde{T}_{ij} = \sum_{s=1}^{D} \sum_{t=1}^{D} L_{ij,st}^2, \quad L_{ij,st} = n^{-1/2} \sum_{k=1}^{n} \frac{(I_{k,i,s} - \nu_s)(I_{k,j,t} - \nu_t)}{(\nu_s \nu_t)^{1/2}}.$$

By multivariate central limit theorem, we can see that $(L_{ij,11},\dots,L_{ij,DD})$ weakly converges to a degenerated multivariate Gaussian distribution. The key is to derive the range of $[0, A_n]$, where the convergence is fast enough compares to the decaying of the multivariate Gaussian tail probability. To this end, we use the Edgeworth expansion to show that $A_n$ could be as large as $o(n^{1/3}D^{-4})$. Apparently, for any $C_0>0$, $C_0\log(n\vee p)$ is within the range $[0, A_n]$. Then we integrate over an eclipse region to show that $\tilde{T}_{ij}$ asymptotically follows $\chi^2\{(D-1)^2\}$ under $\mathrm{H}_{0,ij}$ and the rate of convergence is also fast over the range $[0, A_n]$. On the other hand, when $t\in[0, C_0\log(n\vee p)]$, for any constant $C_0$, the extra term $R_{ij,st}$ can also bounded by the decaying rate $\{\log(n\vee p)\}^{-1}$ with high probability converging to 1. These steps build up the proof of Theorem 1, which is provided in Appendix A.

**Theorem 2.** *Suppose Conditions C1-C3 hold. For any $c_1>0$ and $c_2>2M_1+M_2+c_1$,*

$$\lim_{n\to\infty} P\left\{\inf_{(i,j)\in\mathbb{B}_{c_2}} T_{ij} > c_1\log(n\vee p)\right\} = 1.$$

**Corollary 2.** *Suppose Conditions C1-C3 hold. If for $(i,j)\in\mathbb{H}_1$, $|\gamma_{ij}|>(2M_1+M_2+c_1)\{\log(n\vee p)/n\}^{1/2}$,*

$$\lim_{n\to\infty} P\{T_{ij} > c_1\log(n\vee p)\mid (i,j)\in\mathbb{H}_0\} = 0,$$

$$\lim_{n\to\infty} P\{T_{ij} > c_1\log(n\vee p)\mid (i,j)\in\mathbb{H}_1\} = 1.$$

Although Corollary 2 shows that $T_{ij}$'s null and the alternative distributions are well separated when the alternative quantile association is high, this result is not sufficient for controlling the false discovery rate. In practice, both large and small non-zero quantile associations may exist, and the assumption that the minimal non-zero association is greater than a certain level is often unrealistic. Theorem 3 does not rely on the condition to bound the minimum non-zero association.

**Theorem 3.** *Under Conditions C1-C4, the* FDR *and* FDP *of the multiple testing procedure (6) satisfy* $\lim_{n,p\to\infty}\mathrm{FDR}=\alpha$, *and* $\mathrm{FDP}/\alpha$ *converges to* 1 *in probability as* $n\to\infty$.

One key step to prove Theorem 3 is to show that the dependence among $T_{ij}$ will not affect the validity of the multiple testing procedure. The result relies on some mild condition on the network sparcity (Condition C4). Such condition and the proof technique has been used in other existing literature (*e.g.*, Cai et al. (2013a)).

## 4. Numerical Experiments

We perform extensive numerical experiments to evaluate the performance of the proposed multiple testing procedure. Consider the model

$$Y_i = \beta_{0,i} + \beta_{1,i}X_1 + \beta_{2,i}(U_{0,i})X_2 + \sigma_i Y_{0,i}, \quad Y_{0,i} = F_Y^{-1}(U_{0,i}), \quad i \in \{1, \dots, p\}. \tag{14}$$

We simulate $X_1$ from the truncated Gaussian $\mathrm{TN}(0, 0.2)$ with the range $[-2, 2]$, and $X_2$ from the binary distribution $\mathrm{Bin}(1, 0.3)$. The coefficients $\beta_{0,i}$ and $\beta_{1,i}$ are constants, and $\beta_{2,i}(U_{0,i}) = \beta_{20,i} + U_{0,i}$. We generate $\beta_{0,i}$ from $\mathrm{Unif}(0, 0.5)$, $\beta_{1,i}$ from $0.5\mathrm{N}(0.1, 0.3) + 0.5\mathrm{N}(-0.1, 0.3)$, and $\beta_{20,i}$ from $0.5\mathrm{N}(0.1, 0.3) + 0.5\mathrm{N}(-0.1, 0.3)$. The standard error $\sigma_i$ is a constant, whose value in the simulation is generated from $\mathrm{Unif}(0.2, 0.5)$. Based on (14), the $\tau$-th marginal quantile of $Y_i$ conditioning on $X_1$ and $X_2$ is

$$Q_\tau(Y_i \mid \boldsymbol{X} = (X_1, X_2)') = \beta_{0,i} + \beta_{1,i}X_1 + (\beta_{20,i} + \tau)X_2 + \sigma_i Q_\tau(Y_{0,i}).$$

Here $Q_\tau(Y_{0,i})$ is $\tau$-th marginal quantile of $Y_{0,i}$, whose CDF function is denoted by $F_Y$. We let $F_Y$ take the standard Gaussian distribution under **SE1**–**SE5**, and the standard Cauchy distribution under **SE6**. A brief summary of the simulation settings is provided below, and the detailed descriptions are deferred to Section S3 in the supplementary materials.

**SE1** Linear dependence, Gaussian-tail: The dependence of among $Y_{0,i}$ can be fully captured by its covariance matrix.

**SE2** Linear dependence with outliers, Gaussian-tail: this setting is the same as **S1**, except that the samples are contaminated by 10% outliers generated from the Cauchy distribution.

**SE3** Quadratic dependence, Gaussian-tail: if $Y_{0,i}$ and $Y_{0,j}$ are dependent, their relationship follow a quadratic pattern.

**SE4** Dependence affected by latent variables, Gaussian-tail: If $Y_{0,i}$ and $Y_{0,j}$ is dependent, their association level is affected by some unobserved latent variables.

**SE5** Dependence affected by marginal values of variables, Gaussian-tail. If $Y_{0,i}$ and $Y_{0,j}$ is dependent, their association level is affected by their marginal values.

**SE6** Quadratic dependence, heavy-tail (Cauchy): if $Y_{0,i}$ and $Y_{0,j}$ are dependent, their relationship follow a quadratic pattern.

### *4.1. Numerical Experiments on Testing Conditional Quantile Associations*

In our first set of numerical experiments, we compare the proposed SQUAC method with the methods testing linear association based on data generated by Model (14). For SQUAC, we consider different quantile levels $D = 3$, $D = 4$ and $D = 5$, denoted by SQUAC(3), SQUAC(4) and SQUAC(5) respectively in Table 1. All quantile points $\tau_i$ are chosen evenly based on $D$. We consider different settings of $D$ to illustrate that our method is robust to the choice of $D$. For linear methods, we first regress the covariates $X_1$ and $X_2$ on each $Y_i$ and then get the residuals $R_{k,i}$ for $k = 1, \ldots, n$. Next, we compute standardized residual covariance between each pair,

$$\hat{\xi}_{ij} = \frac{n\hat{\sigma}_{ij}}{\left[\sum_{k=1}^{n}\{(R_{k,i} - \bar{R}_i)(R_{k,j} - \bar{R}_j) - \hat{\sigma}_{ij}\}^2\right]^{1/2}},$$

where $\hat{\sigma}_{ij} = n^{-1}\sum_{k=1}^{n}(R_{k,i} - \bar{R}_i)(R_{k,j} - \bar{R}_j)$, $\bar{R}_i = n^{-1}\sum_{k=1}^{n} R_{k,i}$ and $\bar{R}_j = n^{-1}\sum_{k=1}^{n} R_{k,j}$. We then use the methods proposed in Cai and Liu (2016) to test if $Y_i$ and $Y_j$ is linearly associated conditioning on covariates. The paper discussed two testing methods, one used the asymptotic null distribution of $\hat{\xi}_{ij}$ to perform testing (denoted by LIN-DEP), and the other used a bootstrap method to perform testing (denoted by LIN-DEP-B).

We run 100 repetitions for each simulation setting. We consider two high-dimensional scenarios $(n, p) = (300, 100)$ and $(n, p) = (300, 1000)$. Under all settings, we include 30 dependent pairs among a total of $p(p-1)/2$ pairs. The results are summarized by the average empirical false discovery rates ($\widehat{\text{FDR}}$) and false non-discoveries ($\widehat{\text{FN}}$) across 100 repetitions.

The optimal $\widehat{\text{FDR}}$ is 0.05. Because there are 30 dependent pairs of variables in all settings, the range of $\widehat{\text{FN}}$ is 0 to 30, with 0 being the optimal value.

Table 1 displays the results for all five methods. Under **SE1**, LIN-DEP and LIN-DEP-B outperform the SQUAC by having higher power. Under **SE2**–**SE4** and **SE6**, SQUAC performs satisfactorily while LIN-DEP and LIN-DEP-B can no longer control the FDR and suffer from low power. Under **SE5**, although the linear methods LIN-DEP and LIN-DEP-B can control $\widehat{\text{FDR}}$ under the desired level, their power is much worse than the SQUAC methods. These results confirm the powerful performance of SQUAC in the presence of covariates, outliers and complicated dependence.

Table 1 also shows that the SQUAC method has robust performance with respect to $D$. Both their $\widehat{\text{FDR}}$ and $\widehat{\text{FN}}$ are similar across $D = 3, 4, 5$. For **SE3–SE6**, the false non-discoveries ($\widehat{\text{FN}}$) for $D = 4$ or $D = 5$ are smaller than those for $D = 3$. However, under most settings, the false discovery rates ($\widehat{\text{FDR}}$) levels for $D = 4$ or $D = 5$ are larger than those for $D = 3$ and sometimes slightly inflated (*e.g.*, empirical $\widehat{\text{FDR}} = 0.06$ while the desired level is 0.05). As a common practice in hypothesis testing, the validity of multiple testing is more important than its power. We therefore recommend to use $D = 3$ with a small to moderate sample size.

The SQUAC method is very efficient in computation. For $p = 1000$ and $n = 300$, one repetition of the SQUAC method only takes about 1 minute on a quad-core machine with Intel Core i7-4771 3.5 Hz CPU Processor and 16Gb memory.

*4.2. Numerical Experiments on Testing Marginal Quantile Associations*

Next, we conduct numerical experiments to demonstrate that the SQUAC outperforms other competitive methods because of its capability to measure complex associations. We do not consider covariates here, and compare the SQUAC method with LIN-DEP, LIN-DEP-B, and two other multiple testing methods based on the Kendall's $\tau$ coefficient (KENDALL) and the spearman's $\rho$ coefficient (SPEARMAN). Both methods (KENDALL and SPEARMAN) are newly developed by us to test the rank-associations. The details of these two multiple testing methods are presented in the supplementary materials.

| | Setting 1 | | Setting 2 | | Setting 3 | | Setting 4 | | Setting 5 | | Setting 6 | |
|---|---|---|---|---|---|---|---|---|---|---|---|---|
| | $\widehat{\text{FDR}}$ | $\widehat{\text{FN}}$ | $\widehat{\text{FDR}}$ | $\widehat{\text{FN}}$ | $\widehat{\text{FDR}}$ | $\widehat{\text{FN}}$ | $\widehat{\text{FDR}}$ | $\widehat{\text{FN}}$ | $\widehat{\text{FDR}}$ | $\widehat{\text{FN}}$ | $\widehat{\text{FDR}}$ | $\widehat{\text{FN}}$ |
| | | | | | | $n = 300$, $p = 100$ | | | | | | |
| SQUAC(3) | 0.05 | 6.41 | 0.04 | 9.99 | 0.04 | 0.04 | 0.05 | 9.56 | 0.06 | 0.83 | 0.05 | 0.05 |
| SQUAC(4) | 0.06 | 7.08 | 0.05 | 11.11 | 0.06 | 0.00 | 0.05 | 6.70 | 0.06 | 0.00 | 0.05 | 0.00 |
| SQUAC(5) | 0.07 | 8.07 | 0.06 | 12.66 | 0.06 | 0.00 | 0.06 | 5.87 | 0.06 | 0.00 | 0.05 | 0.00 |
| LIN-DEP-B | 0.05 | 3.53 | 0.16 | 28.26 | 0.97 | 29.97 | 0.99 | 29.99 | 0.05 | 20.09 | 1.00 | 30.00 |
| LIN-DEP | 0.05 | 3.66 | 0.15 | 28.59 | 0.97 | 29.97 | 0.99 | 29.99 | 0.05 | 20.09 | 1.00 | 30.00 |
| | | | | | | $n = 300$, $p = 1000$ | | | | | | |
| SQUAC(3) | 0.05 | 10.91 | 0.05 | 14.90 | 0.05 | 0.42 | 0.04 | 15.14 | 0.05 | 6.47 | 0.04 | 0.45 |
| SQUAC(4) | 0.05 | 11.22 | 0.06 | 15.95 | 0.05 | 0.00 | 0.06 | 10.14 | 0.05 | 0.14 | 0.06 | 0.00 |
| SQUAC(5) | 0.05 | 12.49 | 0.06 | 16.92 | 0.07 | 0.00 | 0.05 | 9.20 | 0.06 | 0.00 | 0.06 | 0.00 |
| LIN-DEP-B | 0.04 | 7.71 | 0.49 | 29.49 | 1.00 | 30.00 | 1.00 | 30.00 | 0.05 | 20.67 | 1.00 | 30.00 |
| LIN-DEP | 0.03 | 7.93 | 0.51 | 29.33 | 1.00 | 30.00 | 1.00 | 30.00 | 0.03 | 20.76 | 1.00 | 30.00 |

Table 1: Simulation results for testing conditional dependence. $\widehat{\text{FDR}}$: the average empirical FDR across 100 repetitions. $\widehat{\text{FN}}$: the average number of false non-discoveries across 100 repetitions.

Table 2 displays the comparison results. Under **SE1**, where the dependence among variables are linear, LIN-DEP and LIN-DEP-B perform the best. The SQUAC has worse $\widehat{\text{FN}}$ than other methods but controls the FDR well. **SE2** evaluates whether the methodas are robust to a moderate percent of outliers. The LIN-DEP and LIN-DEP-B perform unsatisfactorily in this case, and all other methods still work reasonably well. **SE3**–**SE5** examines the performance of these methods when complex association is present. Under both **SE3** and **SE4**, SQUAC is the only method that can still control FDR, while maintaining relatively low $\widehat{\text{FN}}$. The four other methods miss almost all dependent pairs and detected many false signals. Under **SE5**, although all methods can control FDR, they are not as powerful as the SQUAC method. **SE6** simulates heavy tail random variables, where the SQUAC method significantly outperform other methods in both FDR control and power. The results with $p = 1000$ are slightly worse than those with $p = 100$ for all five methods, but the relative performance patterns of these five methods remain the same. From these results, we observe that the SQUAC method is rather robust and can achieve adequate power under various settings, including in the presence of realistic complications such as outliers, non-linear associations, and heavy-tail random variables.

| | Setting 1 | | Setting 2 | | Setting 3 | | Setting 4 | | Setting 5 | | Setting 6 | |
|---|---|---|---|---|---|---|---|---|---|---|---|---|
| | $\widehat{\text{FDR}}$ | $\widehat{\text{FN}}$ | $\widehat{\text{FDR}}$ | $\widehat{\text{FN}}$ | $\widehat{\text{FDR}}$ | $\widehat{\text{FN}}$ | $\widehat{\text{FDR}}$ | $\widehat{\text{FN}}$ | $\widehat{\text{FDR}}$ | $\widehat{\text{FN}}$ | $\widehat{\text{FDR}}$ | $\widehat{\text{FN}}$ |
| $n = 300$, $p = 100$ | | | | | | | | | | | | |
| SQUAC(3) | 0.05 | 6.49 | 0.04 | 10.19 | 0.04 | 0.04 | 0.05 | 8.78 | 0.05 | 0.78 | 0.04 | 0.04 |
| LIN-DEP-B | 0.05 | 1.72 | 0.30 | 28.66 | 0.98 | 29.98 | 0.99 | 29.99 | 0.05 | 20.00 | 1.00 | 30.00 |
| LIN-DEP | 0.05 | 1.76 | 0.28 | 29.07 | 0.98 | 29.98 | 0.99 | 29.99 | 0.05 | 20.00 | 1.00 | 30.00 |
| KENDALL | 0.05 | 1.84 | 0.05 | 5.35 | 0.95 | 29.95 | 0.87 | 29.87 | 0.05 | 19.91 | 0.95 | 29.95 |
| SPEARMAN | 0.05 | 1.84 | 0.05 | 5.35 | 0.96 | 29.96 | 0.91 | 29.91 | 0.06 | 19.96 | 0.96 | 29.96 |
| $n = 300$, $p = 1000$ | | | | | | | | | | | | |
| SQUAC(3) | 0.05 | 10.92 | 0.05 | 15.42 | 0.04 | 0.30 | 0.04 | 12.74 | 0.04 | 6.49 | 0.04 | 0.38 |
| LIN-DEP-B | 0.05 | 4.70 | 0.84 | 29.82 | 1.00 | 30.00 | 1.00 | 30.00 | 0.05 | 20.03 | 1.00 | 30.00 |
| LIN-DEP | 0.04 | 4.84 | 0.84 | 29.72 | 1.00 | 30.00 | 1.00 | 30.00 | 0.04 | 20.03 | 1.00 | 30.00 |
| KENDALL | 0.04 | 4.73 | 0.05 | 10.26 | 1.00 | 30.00 | 0.97 | 29.97 | 0.04 | 20.00 | 1.00 | 30.00 |
| SPEARMAN | 0.05 | 4.69 | 0.06 | 10.19 | 1.00 | 30.00 | 0.99 | 29.99 | 0.06 | 20.00 | 1.00 | 30.00 |

Table 2: Simulation results for testing marginal dependence. $\widehat{\text{FDR}}$: the average empirical FDR across 100 repetitions. $\widehat{\text{FN}}$: the average number of false non-discoveries across 100 repetitions.

We did not compare our methods with other estimates that measures complex associations because their asymptotic null distributions are usually hard to derive. Consequently, it is hard to construct a multiple testing method to control false discovery rate.

## 5. Data Analysis of A Gastric Cancer Study

We illustrate the proposed method using the gastric cancer gene expression data generated by Lee et al. (2014). The full data set can be downloaded from https://www.ncbi.nlm.nih.gov/geo/query/acc.cgi?acc=GSE26253. It contains 432 patients diagnosed with gastric cancer: 68 of them are in stage I, 167 are in stage II, 130 are in stage III, and 67 are in stage IV. Their five-year disease-free survival time and recurrence status are also collected. Previous studies have shown that although gastric cancer survival benefit from adjuvant chemotherapy or chemoradiation therapy (Noh et al., 2014; Bang et al., 2012), the five-year disease-free survival rate remains poor for the patients diagnosed with late stage (III or IV) gastric cancer. In contrast, the patients diagnosed with early stage (I or II) gastric cancer have a better 5-year disease free survival rate (Kim et al., 2005; Lee et al., 2004).

In this analysis, we separate the data set into two, one with early stage cancer patients ($n = 235$), and the other with late stage cancer patients ($n = 197$). We focus on the 85 genes belonging to the Transforming Growth Factor-$\beta$ (TGF-$\beta$) signaling pathway because it plays a complex role in carcinogenesis, having both tumor suppressors and promotors (de Caestecker et al., 2000). The gene identifies and pathway information is available at the KEGG database: `http://www.genome.jp/kegg-bin/show_pathway?hsa04350`. Our goal is to investigate the gene co-expression network for the early-stage and late-stage patients respectively, and further to infer the differential network.

In each dataset (early or late stage), we fit the quantile regression model (1) with the exact stage (I, II, III or IV) as the covariate, and applied the proposed multiple testing procedure (6) to identify conditional association between the pairs of gene expressions. The quantile points are set at $\tau_1 = 0.33$ and $\tau_2 = 0.67$, and the FDR is set at 5%. For each data set, we infer a gene co-expression network of $p$ genes. If the conditional independence is rejected, we draw an edge between the two correspondence genes. Our method identifies 1313 edges in the early stage graph and 1098 edges in the late stage graph. We claim a differential edge if this edge exists in one graph but does not exist in the other. The proposed multiple testing procedure identified 827 differential edges.

As a comparison, we applied the alternative methods to the same data sets. For linear methods (the regression version of LIN-DEP and LIN-DEP-B), we use the exact stage as the covariate. For rank-association-based methods (KENDALL and SPEARMAN), we do not specify any covariate because they cannot adjust for any. The FDR is set at 5%. The numbers of differential edges identified by each method are displayed in Table 3. Our method detected more differential edges than the alternative methods.

One way to check the validity of the detected differential edges is to examine the association between each differential edge and the survival outcome. Specifically, if the edge between $Y_i$ and $Y_j$ is differential, it is likely the joint distribution of $(Y_i, Y_j)$ will affect the survival outcome. To check this point, we combine the early stage patients and late stage patients into one dataset. For subject $k$, $k = 1, 2, ..., 432$, we use an $8-$dimensional indicator vector to summarize the location of his/her paired outcome $(Y_{ki}, Y_{kj})$ in the joint distribution, accounting for disease stage. Specifically, let $\boldsymbol{S}_k = (S_{k1}, S_{k2}, S_{k3})'$ be the dummy variable of the 4 stages of the $k$-th patient. Let $\hat{q}_{is}(\boldsymbol{S}_k)$ $(s = 0, 1, 2, 3;\ i = 1, \ldots, p)$ be the estimated marginal $\tau_s$ quantile of $Y_i$, where $\hat{q}_{i0}(\boldsymbol{S}_k) = -\infty$, and $\hat{q}_{i3}(\boldsymbol{S}_k) = \infty$. We set $J_{k,ij,t}$ as an indicator of the conditional quantile cell that $(Y_{ki}, Y_{kj})$ falls into.

$$J_{k,ij,d} = I\left\{Y_{ki} \in (\hat{q}_{i,s_1-1}(\boldsymbol{S}_k), \hat{q}_{i,s_1}(\boldsymbol{S}_k)],\ Y_{kj} \in (\hat{q}_{j,s_2-1}(\boldsymbol{S}_k), \hat{q}_{j,s_2}(\boldsymbol{S}_k)]\right\}, \quad d = 3(s_1 - 1) + s_2.$$

It is easy to see that $\sum_{d=1}^{9} J_{k,ij,d} = 1$. Let $\boldsymbol{J}_{k,ij} = (J_{k,ij,2}, \ldots, J_{k,ij,9})'$ be the 8-dimensional covariate. If the differential edge between Gene $i$ and Gene $j$ is associated with survival, at least one $J_{k,ij,d}$ $(d = 2, \ldots, 9)$ should be associated with the survival rate. Then, we fit a cox proportional hazard model $\lambda(v \mid \boldsymbol{J}_{k,ij}) = \lambda_{0,ij}(v) \exp(\sum_{d=1}^{8} \beta_{ij,d} J_{k,ij,d+1})$, where $\lambda(v)$ is the hazard function, and $\lambda_{0,ij}(v)$ is the baseline hazard function that both Gene $i$ and Gene $j$ falls into the lowest sample tertile. We set the null hypothesis $H_{0,ij} : \beta_{ij,d} = 0,\ \forall d = 1, \ldots, 8$, and derived its corresponding $p$-values and their estimated density. For each method, we first determine the set of differential edges and then perform the above procedure to get the estimated density curve (presented in Figure 1). It is clear that the $p$-values corresponding to SQUAC method tend to take smaller values, indicating that that the paired outcomes (e.g., edges) detected by SQUAC are more likely to be associated to the survival outcomes.

To further illustrate the application of the proposed method, we count the degree (the number of edges stemming out of one vertex) of each gene in early and late stage gene co-expression networks identified by SQUAC, and calculate the degree differences between two networks. The top 5 genes with the largest degree differences are *TGF-β3*, *BMP4*, *DCN*,

*AMHR2* and *SMAD4*. *TGF-$\beta$3*, as a *TGF-$\beta$* family member, has been discovered to play a dual role in human cancer in different prognostic stages of cancer – in early stages, it performs as a tumor suppressor, while in the late stages, it performs as a tumor promotor (Jakowlew, 2006; Lebrun, 2012). *BMP4* has been identified as a modulator of cisplatin (a widely used gastric cancer chemotherapy) sensitivity in gastric cancer (Ivanova et al., 2013). *DCN* is capable of suppressing the growth of multiple types of tumor. For example, evidence has been shown that it is a key regulator for chemoresistant mechanisms for oral cancer (Kasamatsu et al., 2015), and associated with breast cancer metastasis and survival (Cawthorn et al., 2012; Ishiba et al., 2014). *AMHR2* has been recently shown to regulate survival signaling in non-small cell lung cancer (Beck et al., 2015). *SMAD4* has been shown to be significantly related to the prognostic differences in gastric cancer patients as well as the survival rates (Leng et al., 2009; Wang et al., 2007). This result demonstrates the practical utility of SQUAC in real-life biomedical studies.

| Method | SQUAC | KENDALL | SPEARMAN | LIN-DEP | LIN-DEP-B |
|---|---|---|---|---|---|
| Number | 827 | 727 | 733 | 731 | 66 |

Table 3: Number of differential edges

## 6. Extension to Quantile Regression in High Dimensional Sparse Models

In genetics and genomics study, high dimensional covariates may affect the expression levels. For example, in the expression quantitative trait loci (eQTL) studies, in addition to gene expression levels, thousands or even millions of genetic markers are measured. Researchers are interested in the gene co-expression patterns conditioning on the genetic markers. The null hypotheses are $\mathrm{H}_{0,ij} : Y_i \perp\!\!\!\perp Y_j \mid \boldsymbol{X}$, where $Y_i$ and $Y_j$ are expression levels of Gene $i$ and Gene $j$, and $\boldsymbol{X} = (X_1, \ldots, X_{p_x})'$ are genetic markers. We use the same quantile regression model (1). Since only a few genetic markers will affect the gene expression level of a certain gene, we assume $\boldsymbol{\beta}_{0,i}(\tau_s)$ is a sparse vector for all genes. Let $s_x = \sup_{1\le i\le p, 1\le s\le D-1} \|\boldsymbol{\beta}_{0,i}(\tau_s)\|_0$.

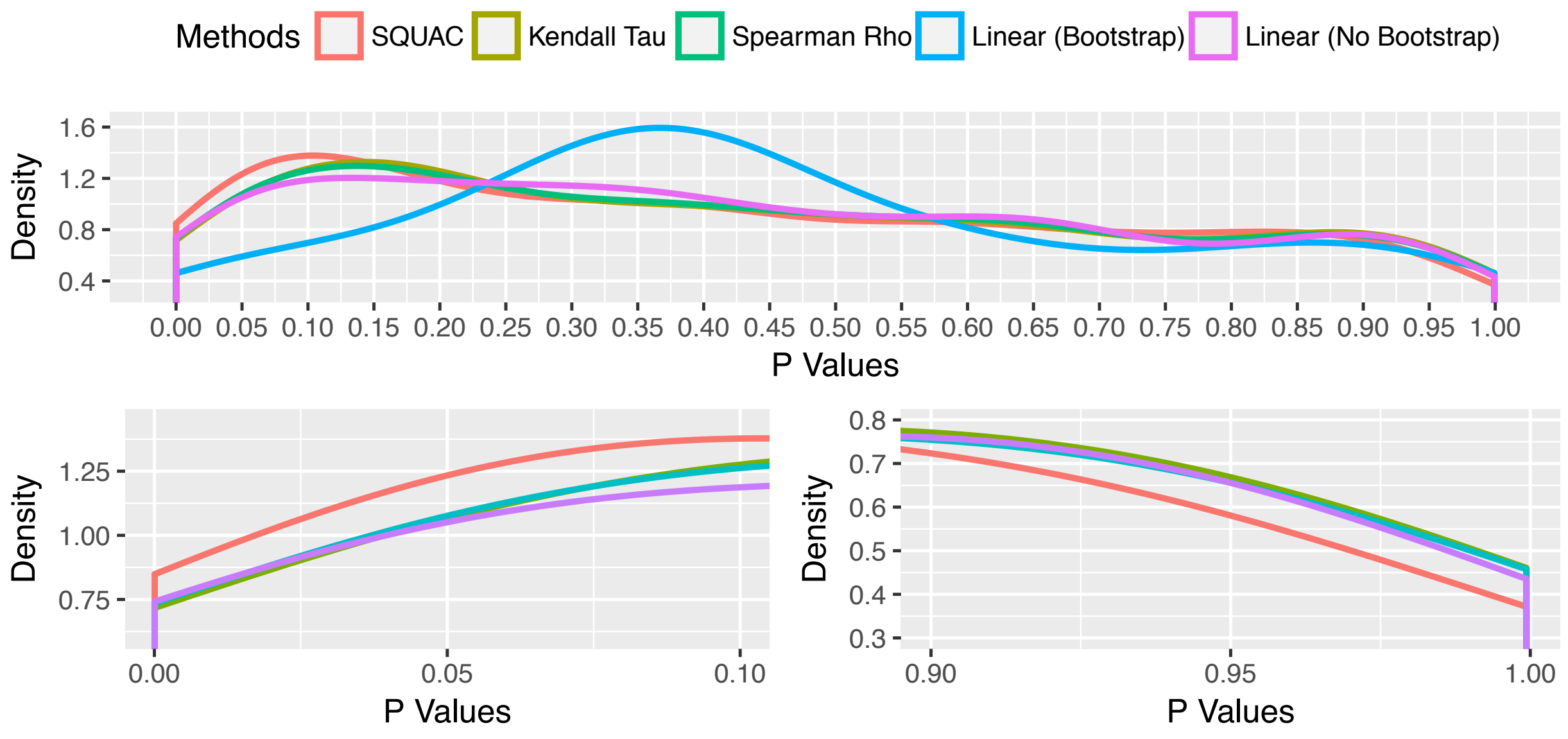

Figure 1: Density histogram of $p$-values after regressing quantile cell indicators on survival. Top: density histogram of $p$-values of each method; bottom left: enlarged density histogram of $p$-values of each method (excluding LIN-DEP-B) in the interval $(0, 0.1)$; bottom right: enlarged density histogram of $p$-values of each method in the interval $(0.9, 1)$.

It represents the maximum sparsity level of all quantile coefficient vectors, which is usually much smaller than $n$.

When sparse high dimensional covariates exist, the traditional coefficient estimation method (2) no longer works; penalized regression methods are usually used to yield consistent estimators. Many papers have discussed such procedures. See, for example, Belloni and Chernozhukov (2011), Wang et al. (2012), Zheng et al. (2013), Fan et al. (2014), and Zheng et al. (2015). These papers also propose different conditions and the convergence rate of quantile levels that can be achieved under these conditions.

Considering these convergence rate under the case $p_x \to \infty$, we propose to adjust the Conditions C2 to make sure the main results hold for the high-dimensional sparse covariates setting.

C2*. Assume $p_x \leq cn^r$ for some $r > 0$, $s_x$ is a constant, and there exists a constant $C_2$ such

that

$$\mathsf{P}\left[\sup_{s=1}^{D}|\hat{Q}_{k,i,s}-Q_{k,i,s}|>2C_2(1+\varepsilon)n^{-1/2}\{\log(n\vee p)\}^{1/2}\right]<D(n\vee p)^{-2-\varepsilon}. \qquad (15)$$

Some existing methods enjoy the property described in (15) under certain conditions, such as the adaptive robust Lasso method developed in Fan et al. (2014). A description on why this method yields estimates satisfying Condition C2* is provided in Section S5 of the supplementary materials.

## 7. Discussion

In this paper, we use quantile associations to measure the general association between variables and develop the SQUAC test statistics to measure the quantile dependence. The proposed testing and multiple testing procedures are based on the SQUAC statistics. Commonly used surrogates for associations, such as linear associations (correlation) and rank associations (Kendall's $\tau$ or Spearman's $\rho$ coefficients), require parametric or semi-parametric assumptions to hold to fully capture the dependence between variables. Quantile associations require considerably weaker assumptions. They require neither assumptions on the parametric form of the distributions nor the association structures. Therefore, they are robust and flexible enough to be applied to measure complicated associations.

When deriving quantile associations, the quantile levels and their number $D$ are predetermined. In genomic applications, it is reasonable to use tertile points to cut gene expression levels because it is easy to interpretate the results. Based on our simulation results, we recommend using $D=3$ for small to moderate sample size. In general, as $D$ increase, the SQUAC statistic can capture more local associations and become more and more powerful to test general associations. On the other hand, large $D$ will slow down the convergence of SQUAC statistics' asymptotic null distribution. For simplicity, we only use even grids of quantile levels for our method. Nevertheless, both the value of $D$ and the quantile points will affect the amount of quantile association captured by the proposed method, and therefore will affect its power. For different pairs of $(Y_i, Y_j)$, the value of $D$ and the quantile levels can be set differently and adaptively to capture weak general associations. Because the distribution of $(Y_1, \ldots, Y_p)'$ is usually unknown, the derivation of optimal $D$ and the quantile levels is not easy. We will investigate the data adaptive method to choose their optimal values in the future.

The current theoretical results of the SQUAC method are based on the linear quantile regression model assumption to obtain appropriate conditional quantile estimates. It is possible to extend the idea to models beyond linear quantile regression model towards more general models, such as nonparametric quantile regression models or even complete non-parametric models. Theoretically, without any model assumption of each $Y_i$ and $\boldsymbol{X}$, one can still estimate the conditional distribution function in a nonparametric framework to get the estimate of the quantiles. As long as Condition C1 and C2 are satisfied, the theoretical results can be extended for more general cases.

### Acknowledgments

The authors are grateful to three referees, the Association Editor, and the Editor for their thoughtful and suggestive comments, which have helped to improve greatly on an earlier manuscript. Jichun Xie's research was partially supported by the NIH CTSA grant UL1TR002553, and Ruosha Li's research was partially supported by NSF DMS grant 1612965.

### Supplementary Materials

We include in the supplementary materials the connection and difference between our method and other quantile association inference methods, the proofs of the lemmas, a detailed description of the six simulation settings, the multiple testing methods KENDALL and SPEARMAN, and a description on why the adaptive Lasso method developed in Fan et al. (2014) yields estimates that satisfy Condition C2*. The supplementary materials are provided in a separate pdf file.

## Appendix A. Proofs of the Theorems

For all $k$, $i$, $s$, define

$$\hat{I}_{k,i,s} = I(\hat{Q}_{k,i,s-1} < Y_{ki} \leq \hat{Q}_{k,i,s}), \quad I_{k,i,s} = I(Q_{k,i,s-1} < Y_{ki} \leq Q_{k,i,s}).$$

An equivalent form of the SQUAC statistic is

$$T_{ij} = \sum_{s=1}^{D}\sum_{t=1}^{D}\left\{n^{-1/2}\sum_{k=1}^{n}\frac{\hat{I}_{k,i,s}\hat{I}_{k,j,t} - \nu_s\nu_t}{(\nu_s\nu_t)^{1/2}}\right\}^2.$$

Lemmas 1 – 7 are required to prove the asymptotic properties of the proposed method. Among those, Lemmas 1 – 5 are listed below, and Lemmas 6 – 7 are listed in the proof of the theorems.

**Lemma 1.**

$$\Delta_{ij,st} = n^{-1/2}\sum_{k=1}^{n}\frac{\hat{I}_{k,i,s}\hat{I}_{k,j,t} - \nu_s\nu_t}{(\nu_s\nu_t)^{1/2}} - n^{-1/2}\sum_{k=1}^{n}\frac{(I_{k,i,s} - \nu_s)(I_{k,j,t} - \nu_t)}{(\nu_s\nu_t)^{1/2}}. \tag{A.1}$$

*Then for all* $\varepsilon > 0$,

$$P\left[\sup_{(i,j)\in\mathbb{H}}\sup_{1\leq s,t\leq D}|\Delta_{ij,st}| > (M_2+\varepsilon)\{\log(n\vee p)\}^{1/2}\right] \leq CD(n\vee p)^{-\varepsilon/M}. \tag{A.2}$$

*with* $M_2 = (3-2u_0)(4C_1C_2+2)/u_0$ *and some sufficiently large* $M$.

*Further, depending on positive integer* $m$,

$$\sup_{(i,j)\in\mathbb{H}_0}\sup_{1\leq s,t\leq D} E(\Delta_{ij,st}) \leq Cn^{-1/2}, \tag{A.3}$$

$$\sup_{(i,j)\in\mathbb{H}_0}\sup_{1\leq s,t\leq D} E\{|\Delta_{ij,st}|^m\} \leq C_m[\{\log(n\vee p)\}^2/n]^{m/2}., \tag{A.4}$$

*where* $C_m = \{4(1+\varepsilon)C_1C_2\}^m\mu_0^{-m}m!$.

**Lemma 2.** *Suppose* $X$ *follows a chi-square distribution with degree of freedom* $K$*, where* $K$ *is a fixed positive integer. Then*

$$\lim_{t\to\infty}\frac{P(X>t)}{\{\Gamma(K/2)\}^{-1}(t/2)^{K/2-1}e^{-t/2}} = 1. \tag{A.5}$$

**Lemma 3.** *Consider $\mathbb{H}_{01}$ defined in (S28). Under Conditions C1–C4, for any $1 \le t \le b_p$,*

$$\sup_{(a,b,c,d)\in\mathbb{H}_{01}} |\mathsf{P}(T_{ab} > t, T_{cd} > t) - \mathsf{P}(T_{ab} > t)\mathsf{P}(T_{cd} > t)| = o[\{G_D(t)\}^2]. \tag{A.6}$$

**Lemma 4.** *Let $\tilde{T}_{ij} = \sum_{s=1}^{D}\sum_{t=1}^{D}\left\{n^{-1/2}\sum_{k=1}^{n}\frac{(I_{k,i,s}-\nu_s)(I_{k,j,t}-\nu_t)}{(\nu_s\nu_t)^{1/2}}\right\}^2$. Then*

$$\sup_{(i,j)\in\mathbb{H}_0}\left|\frac{\mathsf{P}(\tilde{T}_{ij}\ge t)}{G_D(t)} - 1\right| \le CD^6(1+t)^{3/2}n^{-1/2}, \quad \forall\, t \text{ satisfying } t = o\{n^{1/3}D^{-4}\}. \tag{A.7}$$

**Lemma 5.** *For any positive integer $D$, let $\phi_{I_D}$ be the density of the $D$-dimensional standard multivariate Gaussian distribution, and $Q_K(u)$ be a $K$-th order polynomial of $u \in \mathbb{R}^D$ with bounded coefficients. Then for any $t \ge 0$, $|\int_{\|u\|^2\ge t} Q_K(u)\phi_{I_D}(u)\,\mathrm{d}u| \le C_K D^K(1+t)^{K/2}\int_t^\infty f_{\chi^2_D}(x)\,\mathrm{d}x$, where $C_K$ is a constant only depending on $K$.*

*Proof of Theorem 1.* Let

$$L_{ij,st} = n^{-1/2}\sum_{k=1}^{n}\frac{(I_{k,i,s}-\nu_s)(I_{k,j,t}-\nu_t)}{(\nu_s\nu_t)^{1/2}}. \tag{A.8}$$

Recall $\Delta_{ij,st}$ defined in (A.1). Then $T_{ij} = \tilde{T}_{ij} + R_{ij}$, where

$$\tilde{T}_{ij} = \sum_{s=1}^{D}\sum_{t=1}^{D} L_{ij,st}^2, \quad R_{ij} = 2\sum_{s=1}^{D}\sum_{t=1}^{D} L_{ij,st}\Delta_{ij,st} + \sum_{s=1}^{D}\sum_{t=1}^{D}\Delta_{ij,st}^2. \tag{A.9}$$

When $j_1 \ne j_2$, $\mathsf{E}(I_{k,i,j_1}I_{k,i,j_2}) = 0$, and when $j_1 = j_2$, $\mathsf{E}(I_{k,i,j_1}I_{k,i,j_2}) = \mathsf{E}(I_{k,i,j_1}) = \nu_{j_1}$. Then, for any two pairs $(s_1,t_1)$ and $(s_2,t_2)$, it is easy to see that

$$\mathsf{Cov}(L_{ij,s_1,t_1}, L_{ij,s_2,t_2}) = \begin{cases} (1-\nu_s)(1-\nu_t), & \text{if } s_1 = s_2 = s,\ t_1 = t_2 = t; \\ -(1-\nu_s)(\nu_{t_1}\nu_{t_2})^{1/2}, & \text{if } s_1 = s_2 = s,\ t_1 \ne t_2; \\ -(\nu_{s_1}\nu_{s_2})^{1/2}(1-\nu_t), & \text{if } s_1 \ne s_2,\ t_1 = t_2 = t; \\ (\nu_{s_1}\nu_{s_2})^{1/2}(\nu_{t_1}\nu_{t_2})^{1/2}, & \text{if } s_1 \ne s_2,\ t_1 \ne t_2. \end{cases} \tag{A.10}$$

Define random vector $\boldsymbol{L}_{ij} = (L_{ij,11}, L_{ij,12}\ldots, L_{ij,DD})'$. By (A.10), $\mathsf{Var}(\boldsymbol{L}_{ij}) = \boldsymbol{\Sigma} = \boldsymbol{\Sigma}_1 \otimes \boldsymbol{\Sigma}_1$, where $\boldsymbol{\Sigma}_1 = \mathbf{I}_D - \sqrt{\boldsymbol{\nu}}\sqrt{\boldsymbol{\nu}}'$, with $\sqrt{\boldsymbol{\nu}} = (\sqrt{\nu_1},\ldots,\sqrt{\nu_D})'$. By the Central Limit Theorem, $\boldsymbol{L}$ asymptotically follows the multivariate normal distribution $\mathrm{N}_{D^2}(\mathbf{0}, \boldsymbol{\Sigma}_1 \otimes \boldsymbol{\Sigma}_1)$.

Let $\boldsymbol{\Sigma}_1 = (\sigma_{ij})_{D\times D}$ and $\boldsymbol{\Sigma} = \boldsymbol{\Sigma}_1 \otimes \boldsymbol{\Sigma}_1$. Because $\sqrt{\boldsymbol{\nu}}'\sqrt{\boldsymbol{\nu}} = \sum_{i=1}^{D} \nu_i = 1$, $\boldsymbol{\Sigma}_1$ is idempotent, *i.e.*, $\sum_{k=1}^{D} \sigma_{ik}\sigma_{kj} = \sigma_{ij}$. Then $\boldsymbol{\Sigma}$ is also idempotent, because

$$\boldsymbol{\Sigma}^2 = (\sigma_{ij}\boldsymbol{\Sigma}_1)_{D^2\times D^2} \times (\sigma_{ij}\boldsymbol{\Sigma}_1)_{D^2\times D^2} = \left(\sum_{k=1}^{D} \sigma_{ik}\sigma_{kj}\boldsymbol{\Sigma}_1^2\right)_{D^2\times D^2} = (\sigma_{ij}\boldsymbol{\Sigma}_1)_{D^2\times D^2} = \boldsymbol{\Sigma}.$$

Because $\boldsymbol{\Sigma}_1$ is idempotent and symmetric, $\boldsymbol{\Sigma}_1$ is a projection matrix with $\mathsf{Rank}(\boldsymbol{\Sigma}_1) = \mathsf{tr}(\boldsymbol{\Sigma}_1) = \sum_{i=1}^{D}(1-\nu_i) = D-1$. Therefore, $\mathsf{Rank}(\boldsymbol{\Sigma}) = \{\mathsf{Rank}(\boldsymbol{\Sigma}_1)\}^2 = (D-1)^2$.

By Theorem 5.5A in Rencher (2000), because $\boldsymbol{\Sigma}$ is idempotent with degree of freedom $(D-1)^2$, $\tilde{T}_{ij} = \boldsymbol{L}_{ij}'\boldsymbol{L}_{ij}$ asymptotically follows the central chi-square distribution with the degree of freedom $(D-1)^2$.

Then

$$\begin{aligned}
&\sup_{0\le t\le C_0\log(n\vee p)} \sup_{(i,j)\in\mathbb{H}_0} \frac{\mathsf{P}(T_{ij}>t)}{G_D(t)} \\
\le &\sup_{0\le t\le C_0\log(n\vee p)} \sup_{(i,j)\in\mathbb{H}_0} \left\{\frac{\mathsf{P}(\tilde{T}_{ij}\ge t-\{\log(n\vee p)\}^{-1}) + \mathsf{P}(R_{ij}>\{\log(n\vee p)\}^{-1})}{G_D(t)}\right\} \\
= &\sup_{0\le t\le C_0\log(n\vee p)} \sup_{(i,j)\in\mathbb{H}_0} \left\{\frac{G_D(t-\{\log(n\vee p)\}^{-1})}{G_D(t)} \frac{\mathsf{P}(\tilde{T}_{ij}>t-\{\log(n\vee p)\}^{-1})}{G_D(t-\{\log(n\vee p)\}^{-1})}\right. \\
&\left. + \frac{\mathsf{P}(R_{ij}>\{\log(n\vee p)\}^{-1})}{G_D(t)}\right\} \qquad \text{(A.11)}
\end{aligned}$$

By Lemma 2 and Lemma 4,

When $t$ is a constant, based on the continuity of $G_D(t)$, for any small $\varepsilon_p$, when $p$ sufficiently large,

$$|G_D(t-\{\log(n\vee p)\}^{-1})/G_D(t) - 1| \le \varepsilon_p. \qquad \text{(A.12)}$$

When $t = O\{\log(n\vee p)\}$, Lemma 2 leads to (A.12).

Lemma 4 leads to

$$\left|\frac{\mathsf{P}(\tilde{T}_{ij}>t-\{\log(n\vee p)\}^{-1})}{G_D(t-\{\log(n\vee p)\}^{-1})} - 1\right| \le CD^6\{\log(n\vee p)\}^{3/2}n^{-1/2} \le \varepsilon_p.$$

To prove the final result, we also needs the following lemma.

**Lemma 6.** *When* $0 \le t \le C_0\log(n\vee p)$, $\mathsf{P}(|R_{ij}| > \{\log(n\vee p)\}^{-1})/G_D(t) \le \varepsilon_p$.

Thus,

$$\sup_{0\le t\le C_0\log(n\vee p)}\sup_{(i,j)\in\mathbb{H}_0}\frac{\mathsf{P}(T_{ij}>t)}{G_D(t)}\le 1+3\varepsilon_p.$$

Similarly, by

$$\sup_{0\le t\le C_0\log(n\vee p)}\sup_{(i,j)\in\mathbb{H}_0}\frac{\mathsf{P}(T_{ij}>t)}{G_D(t)}\ge \sup_{(i,j)\in\mathbb{H}_0}\frac{\mathsf{P}(\tilde{T}_{ij}\le t+c)-\mathsf{P}(R_{ij}>c)}{G_D(t)},$$

we can show

$$\sup_{0\le t\le C_0\log(n\vee p)}\sup_{(i,j)\in\mathbb{H}_0}\frac{\mathsf{P}(T_{ij}>t)}{G_D(t)}\ge 1-3\varepsilon_p.$$

This completes the proof. □

*Proof of Theorem 2.* It is easy to see that

$$\sup_{1\le i,j\le p}\sup_{1\le s,t\le D}\left|\frac{(I_{k,i,s}-\nu_s)(I_{k,j,t}-\nu_t)}{(\nu_s\nu_t)^{1/2}}-\gamma_{ij,st}\right|\le M_1,$$

by Azuma's inequality,

$$\mathsf{P}\left[\sup_{1\le s,t\le D}\left|L_{ij,st}-n^{1/2}\gamma_{ij,st}\right|>(2+\varepsilon)M_1\{\log(n\vee p)\}^{1/2}\right]\le 2D^2(n\vee p)^{-(2+\varepsilon)}.$$

Because $\mathsf{Card}(\mathbb{B}_{c_2})\le p^2$,

$$\mathsf{P}\left[\sup_{(i,j)\in\mathbb{B}_{c_2}}\sup_{1\le s,t\le D}\left|L_{ij,st}-n^{1/2}\gamma_{ij,st}\right|>(2+\varepsilon)M_1\{\log(n\vee p)\}^{1/2}\right]\le 2D^2(n\vee p)^{-\varepsilon}.$$

$T_{ij}=\sum_{s=1}^D\sum_{t=1}^D(n^{1/2}\gamma_{ij,st}+L_{ij,st}-n^{1/2}\gamma_{ij,st}+\Delta_{ij,st})^2$. When $|\gamma_{ij,st}|\ge c_2\{\log(n\vee p)/n\}^{1/2}$ and $c_1=c_2-2M_1-M_2-\varepsilon>0$,

$$\begin{aligned}
&\mathsf{P}\left\{\inf_{(i,j)\in\mathbb{B}_{c_2}}T_{ij}>c_1\log(n\vee p)\right\}\\
&\ge 1-\mathsf{P}\left[\sup_{(i,j)\in\mathbb{B}_{c_2}}\sup_{1\le s,t\le D}|\Delta_{ij,st}|>(M_2+\varepsilon/2)\{\log(n\vee p)\}^{1/2}\right]\\
&\qquad-\mathsf{P}\left[\sup_{(i,j)\in\mathbb{B}_{c_2}}\sup_{1\le s,t\le D}\left|L_{ij,st}-n^{1/2}\gamma_{ij,st}\right|\ge\{2+\varepsilon/(2M_1)\}M_1\{\log(n\vee p)\}^{1/2}\right]\\
&\ge 1-CD^2(n\vee p)^{-\varepsilon/M}.
\end{aligned}$$

□

*Proof of Theorem 3.* By the continuity of $G_D(t)$ and the monotonicity of $\sum_{i,j} I(T_{ij} > t)$, we can obtain that for some $0 \le \hat{t} \le t_p$, $G(\hat{t})q/\left\{\max\{\sum_{1\le i<j\le p} I(T_{ij} > \hat{t}), 1\}\right\} = \alpha$.

By Theorem 2, $\sum_{i,j} I\{T_{ij} > t_p\} \ge c_p$. Let $b_p = t_p - 2\log c_p$. By definition of $\hat{t}$, we have $\mathsf{P}(0 \le \hat{t} \le b_p) \to 1$.

To prove this theorem, it suffices to show that

$$\sup_{0\le t\le b_p} \left| \frac{\sum_{(i,j)\in\mathbb{H}_0} I(T_{ij} > t)}{qG_D(t)} - 1 \right| \to 0 \text{ in probability.}$$

Let $e_p(t) = \mathsf{E}\Big[\sum_{(i,j)\in\mathbb{H}_0} I(T_{ij} > t)/\{qG_D(t)\}\Big] = q_0\mathsf{P}(T_{ij} > t)/\{qG_D(t)\}$. By Lemma 1, for $0 \le t \le b_p$, $e_p(t) \to 1$. By Chebyshev's Inequality, $\forall \epsilon > 0$,

$$\mathsf{P}\left(\left|\frac{\sum_{(i,j)\in\mathbb{H}_0} I(T_{ij} > t)}{qG_D(t)} - 1\right| > \epsilon\right) \le \frac{\mathsf{Var}[\sum_{(i,j)\in\mathbb{H}_0} I(T_{ij} > t)/\{qG_D(t)\}] + (e_p(t) - 1)^2}{\epsilon^2}.$$

Therefore, it suffices to show that $\sup_{0\le t\le b_p} \mathsf{Var}[\sum_{(i,j)\in\mathbb{H}_0} I(T_{ij} > t)/(qG_D(t))] \to 0$.

$$\mathsf{Var}\left\{\frac{\sum_{(i,j)\in\mathbb{H}_0} I(T_{ij} > t)}{qG_D(t)}\right\} = \frac{1}{q^2\{G_D(t)\}^2} \sum_{(i,j)\in\mathbb{H}_0} \mathsf{P}(T_{ij} > t)\mathsf{P}(T_{ij} \le t)$$
$$+ \frac{1}{q^2\{G_D(t)\}^2} \sum_{(a,b),(c,d)\in\mathbb{H}_0,(a,b)\ne(c,d)} \{\mathsf{P}(T_{ab} > t, T_{cd} > t) - \mathsf{P}(T_{ab} > t)\mathsf{P}(T_{cd} > t)\}$$

As $p$ sufficiently large, the first term

$$\frac{1}{q^2\{G_D(t)\}^2} \sum_{(i,j)\in\mathbb{H}_0} \mathsf{P}(T_{ij} > t)\mathsf{P}(T_{ij} \le t) \le \frac{2}{\alpha c_p} \frac{\mathsf{P}(T_{ij} > t)}{G_D(t)} \le C c_p^{-1} \to 0.$$

Combined with Lemma 7, we prove the result.

**Lemma 7.** *As $n \vee p \to \infty$,*

$$\frac{1}{q^2\{G_D(t)\}^2} \sum_{(a,b,c,d)\in\mathbb{H}_{02}} \{P(T_{ab} > t, T_{cd} > t) - P(T_{ab} > t)P(T_{cd} > t)\} \to 0.$$

□

## References

The list also includes the reference cited in the supplementary materials. The supplementary materials are available at https://github.com/jichunxie/squac.

# Supplementary Materials for "False Discovery Rate Control for High-Dimensional Networks of Quantile Associations Conditioning on Covariates"

## S1. Connection and Difference between Our method and Other Quantile Association Inference Methods

The idea to measure general associations by quantile associations has been investigated before. It can be traced back to Blomqvist (1950), where one pair of continuous variables are dichotomized, and a measure of their correlation was proposed and investigated. Later, Borkowf et al. (1997) proposed to use agreement test to study the association pattern between one pair of bivariate continous data by categorizing them based on empirical quantiles. Wei (2008) visualized covariate-specific bivariate quantile contours. Recently, Li et al. (2014b) proposed to use quantile-specific odds ratio as a statistic to measure the level of association conditioning on covariates. They propose a statistic to summarize the conditional quantile associations over a range and used an iterative smoothing technique to estimate their null distributions. For high dimensional network inference, such smoothing-based inference procedures for all pairs of outcomes is not computationally feasible. In this work, we propose a test statistic that asymptotically follows chi-square distribution under the null. With the known asymptotic null distribution, the proposed FDR control procedure is very efficient so that it can be easily applied to the application problems with large $p$.

## S2. Proof of the Lemmas

*Proof of Lemma 1.* By Condition C2,

$$\mathsf{P}\left\{\sup_{i=1}^{p}\sup_{k=1}^{n}\sup_{s=1}^{D}|\hat{Q}_{k,i,s}-Q_{k,i,s}|>2(1+\varepsilon)C_2n^{-1/2}\{\log(n\vee p)\}^{1/2}\right\}<D(n\vee p)^{-\varepsilon}.$$

Define

$$\mathcal{X}=\{(\boldsymbol{X}_k,\boldsymbol{Y}_k)_{k=1}^{n}:\sup_{i=1}^{p}\sup_{k=1}^{n}\sup_{s=1}^{D}|\hat{Q}_{k,i,s}-Q_{k,i,s}|\leq 2(1+\varepsilon)C_2n^{-1/2}\{\log(n\vee p)\}^{1/2}\}.$$

By Theorem 2.2 in Koenker (2005),

$$\nu_s n - p_x \leq \sum_{k=1}^{n} \hat{I}_{k,1,s} \leq \nu_s n + p_x, \quad \nu_t n - p_x \leq \sum_{k=1}^{n} \hat{I}_{k,2,t} \leq \nu_t n + p_x \tag{S1}$$

Therefore, $\Delta_{ij,st} = \sum_{l=1}^{3} \Delta_{l,ij,st} + O(n^{-1/2})$, where

$$\Delta_{1,ij,st} = n^{-1/2} \sum_{k=1}^{n} \frac{(I_{k,i,s} - \nu_s)(\hat{I}_{k,j,t} - I_{k,j,t})}{(\nu_s \nu_t)^{1/2}} \tag{S2}$$

$$\Delta_{2,ij,st} = n^{-1/2} \sum_{k=1}^{n} \frac{(I_{k,j,t} - \nu_t)(\hat{I}_{k,i,s} - I_{k,i,s})}{(\nu_s \nu_t)^{1/2}} \tag{S3}$$

$$\Delta_{3,ij,st} = n^{-1/2} \sum_{k=1}^{n} \frac{(\hat{I}_{k,i,s} - I_{k,i,s})(\hat{I}_{k,j,t} - I_{k,j,t})}{(\nu_s \nu_t)^{1/2}} \tag{S4}$$

On the space $\mathcal{X}$, for sufficiently large $n$ and $p$,

$$|\hat{I}_{k,j,t} - I_{k,j,t}| \leq I\{Y_{kj} \text{ is between } Q_{k,i,t} - \delta_{n,p} \text{ and } Q_{k,i,t} + \delta_{n,p}\}. \tag{S5}$$

Then

$$\begin{aligned} |\Delta_{1,ij,st}| &\leq \frac{1-u_0}{u_0} n^{-1/2} \sum_{k=1}^{n} |\hat{I}_{k,j,t} - I_{k,j,t}| \\ &\leq \frac{1-u_0}{u_0} n^{-1/2} \sum_{k=1}^{n} I\{Y_{kj} \text{ is between } Q_{k,i,t} - \delta_{n,p} \text{ and } Q_{k,i,t} + \delta_{n,p}\}, \end{aligned} \tag{S6}$$

for sufficiently large $n$ and $p$. Here, $\delta_{n,p} = 2(1+\varepsilon) C_2 n^{-1/2} \{\log(n \vee p)\}^{1/2}$,

$$\begin{aligned} \mathsf{E}|(S6)| \leq \frac{1-u_0}{u_0} n^{1/2} \mathsf{P}\{Y_{kj} \text{ is between } \boldsymbol{X}_k' \boldsymbol{\beta}_{i0}(\tau_t) \pm \delta_{n,p}\} \\ \leq 4(1+\varepsilon) \frac{1-u_0}{u_0} C_1 C_2 \{\log(n \vee p)\}^{1/2}. \end{aligned}$$

By Azuma's inequality,

$$\mathsf{P}\left[|(S6) - \mathsf{E}\{(S6)\}| > (2+\varepsilon) \frac{1-u_0}{u_0} \{\log(n \vee p)\}^{1/2}\right] \leq 2(n \vee p)^{-(2+\varepsilon)}.$$

It follows that

$$\mathsf{P}\left[\sup_{(i,j) \in \mathbb{H}} \sup_{1 \leq s,t \leq D} |\Delta_{1,ij,st}| > C_{4,\varepsilon} \{\log(n \vee p)\}^{1/2}\right] \leq CD(n \vee p)^{-\varepsilon/M}, \tag{S7}$$

with sufficiently large $M$ and $C_{4,\varepsilon} = \frac{1-u_0}{u_0}(4C_1C_2+2)+\varepsilon$.

Following similar arguments, we can show that

$$\mathsf{P}\left[\sup_{(i,j)\in\mathbb{H}}\sup_{1\le s,t\le D}|\Delta_{2,ij,st}| > C_{4,\varepsilon}\{\log(n\vee p)\}^{1/2}\right] \le CD(n\vee p)^{-\varepsilon/M}, \tag{S8}$$

$$\mathsf{P}\left[\sup_{(i,j)\in\mathbb{H}}\sup_{1\le s,t\le D}|\Delta_{3,ij,st}| > C_{5,\varepsilon}\{\log(n\vee p)\}^{1/2}\right] \le CD(n\vee p)^{-\varepsilon/M}. \tag{S9}$$

Here, $C_{5,\varepsilon} = \frac{1}{u_0}(4C_1C_2+2)+\varepsilon$.

Then, (S2)-(S9) lead to (A.2).

We now prove (A.3) and (A.4).

Under $\mathrm{H}_{0,ij}: Y_i \perp\!\!\!\perp Y_j$, by (S1),

$$\mathsf{E}(\Delta_{ij,st}) = n^{1/2}\frac{\hat{\nu}_{i,s}\hat{\nu}_{j,t}-\nu_s\nu_t}{(\nu_s\nu_t)^{1/2}} \le Cn^{-1/2}.$$

Because $\mathsf{E}|\Delta^m_{ij,st}| \le m!\sum_{l=1}^3 \mathsf{E}|\Delta^m_{l,ij,st}| + m!n^{-m/2}$, to prove (A.4), it suffices to show that

$$\mathsf{E}\Delta^m_{l,ij,st} \le m!\{\log(n\vee p)/n\}^{m/2},\ l=1,2, \quad \mathsf{E}\Delta^m_{3,ij,st} \le m![\{\log(n\vee p)\}^2/n]^{m/2}.$$

$$\begin{aligned}
\mathsf{E}\Delta^m_{1,ij,st} &= \mathsf{E}\left\{n^{-1/2}\sum_{k=1}^n\frac{(I_{k,i,s}-\nu_s)(I_{k,j,t}-\nu_t)}{(\nu_s\nu_t)^{1/2}}\right\}^m \\
&= n^{-m/2}\cdot\mathsf{E}\left\{\sum_{k=1}^n\frac{(I_{k,i,s}-\nu_s)(\hat{I}_{k,j,t}-I_{k,j,t})}{(\nu_s\nu_t)^{1/2}}\right\}^m \\
&= n^{-m/2}(\nu_s\nu_t)^{-m/2}\sum_{m_1+\ldots+m_n=m}\frac{m!}{m_1!\cdots m_n!}\mathsf{E}\left[\prod_{k=1}^n\left\{(I_{k,i,s}-\nu_s)(\hat{I}_{k,j,t}-I_{k,j,t})\right\}^{m_k}\right] \\
&= n^{-m/2}(\nu_s\nu_t)^{-m/2}\sum_{m_1+\ldots+m_n=m}\frac{m!}{m_1!\cdots m_n!}\prod_{k=1}^n\mathsf{E}\left\{(I_{k,i,s}-\nu_s)(\hat{I}_{k,j,t}-I_{k,j,t})\right\}^{m_k}
\end{aligned}$$

When $m_k=0$, $\mathsf{E}\left\{(I_{k,i,s}-\nu_s)(\hat{I}_{k,j,t}-I_{k,j,t})\right\}^{m_k}=1$. We only need to consider those positive $m_k$. Now let $\mathcal{M}_L=\{(m_1,\ldots,m_n):\sum_{k=1}^n I(m_k\ge 1)=L\}$. For any $(m_1,\ldots,m_n)\in\mathcal{M}_L$, let $(k_1,\ldots,k_L)=\{k:m_k\ge 1\}$. Then for any positive even integer $m$ (constant),

$$\begin{aligned}
&\sum_{m_1+\ldots+m_n=m}\frac{m!}{m_1!\cdots m_n!}\prod_{k=1}^n\mathsf{E}\left\{(I_{k,i,s}-\nu_s)(\hat{I}_{k,j,t}-I_{k,j,t})\right\}^{m_k} \\
&\qquad = \sum_{L\in\{1,\ldots,m\}}\sum_{(m_1,\ldots,m_n)\in\mathcal{M}_L}\frac{m!}{m_{k_1}!\cdots m_{k_L}!}\prod_{l=1}^L\mathsf{E}\left\{(I_{k_l,i,s}-\nu_s)(\hat{I}_{k_l,j,t}-I_{k_l,j,t})\right\}^{m_{k_l}}
\end{aligned}$$

When $(i,j)\in\mathbb{H}_0$,

$$\mathsf{E}\left\{(I_{k_l,i,s}-\nu_s)(\hat{I}_{k_l,j,t}-I_{k_l,j,t})\right\}^{m_{k_l}}=\mathsf{E}\left(I_{k_l,i,s}-\nu_s\right)^{m_{k_l}}\mathsf{E}\left(\hat{I}_{k_l,j,t}-I_{k_l,j,t}\right)^{m_{k_l}}.$$

Thus, when $(i,j)\in\mathbb{H}_0$, for any positive even integer $m$,

$$\begin{aligned}\mathsf{E}\Delta_{1,ij,st}^m &\leq n^{-m/2}(\nu_s\nu_t)^{-m/2}\\ &\times\sum_{L\in\{1,\ldots,m\}}\sum_{(m_1,\ldots,m_n)\in\mathcal{M}_L}\frac{m!}{m_{k_1}!\cdots m_{k_l}!}\prod_{l=1}^{L}\left|\mathsf{E}\left(I_{k_l,i,s}-\nu_s\right)^{m_{k_l}}\mathsf{E}\left(\hat{I}_{k_l,j,t}-I_{k_l,j,t}\right)^{m_{k_l}}\right|. \quad \text{(S10)}\end{aligned}$$

Further, when $L>m/2$, there must exist some $m_{k_l}=1$, such that $\mathsf{E}\left(I_{k_l,i,s}-\nu_s\right)^{m_{k_l}}=0$; otherwise, $\sum_{k=1}^n m_k=\sum_{l=1}^L m_{k_l}\geq 2L>m$, contradicting to $\sum_{k=1}^n m_k=m$. By $\mathsf{E}\left(I_{k_l,i,s}-\nu_s\right)=0$, when $L>m/2$,

$$\prod_{l=1}^{L}\left|\mathsf{E}\left(I_{k_l,i,s}-\nu_s\right)^{m_{k_l}}\mathsf{E}\left(\hat{I}_{k_l,j,t}-I_{k_l,j,t}\right)^{m_{k_l}}\right|=0. \quad \text{(S11)}$$

When $L\leq m/2$, by (S5) and $\mathsf{E}\left(I_{k_l,i,s}-\nu_s\right)^{m_{k_l}}\leq 1$,

$$\sup_{1\leq i,j\leq p}\sup_{1\leq s,t\leq D}\prod_{l=1}^{L}\left|\mathsf{E}\left(I_{k_l,i,s}-\nu_s\right)^{m_{k_l}}\mathsf{E}\left(\hat{I}_{k_l,j,t}-I_{k_l,j,t}\right)^{m_{k_l}}\right|\leq(2C_1\delta_{n,p})^L+(n\vee p)^{-\varepsilon}. \quad \text{(S12)}$$

By (S10)-(S12), for postive even number $m$, we have

$$\begin{aligned}&\sup_{1\leq i,j\leq p}\sup_{1\leq s,t\leq D}\mathsf{E}\{|\Delta_{1,ij,st}|^m\}\\ &\leq\{4(1+\varepsilon)C_1C_2\}^{m/2}\mu_0^{-m}m!n^{-m/2}n^{m/2}\{\log(n\vee p)/n\}^{m/2}\leq C_m\{\log(n\vee p)/n\}^{m/2}, \quad \text{(S13)}\end{aligned}$$

where $C_m=\{4(1+\varepsilon)C_1C_2\}^m\mu_0^{-m}m!$.

Similarly, the third term

$$\begin{aligned}\mathsf{E}\Delta_{3,ij,st}^m &\leq n^{-m/2}(\nu_s\nu_t)^{-m/2}\\ &\times\sum_{l\in\{1,\ldots,L\}}\sum_{(m_1,\ldots,m_n)\in\mathcal{M}_L}\frac{m!}{m_{k_1}!\cdots m_{k_l}!}\prod_{l=1}^{L}\left|\mathsf{E}\left(\hat{I}_{k_l,i,s}-I_{k_l,i,s}\right)^{m_{k_l}}\mathsf{E}\left(\hat{I}_{k_l,j,t}-I_{k_l,j,t}\right)^{m_{k_l}}\right|.\end{aligned}$$

Because for $(m_1,\ldots,m_n)\in\mathcal{M}_L$

$$\sup_{1\leq i,j\leq p}\sup_{1\leq s,t\leq D}\prod_{l=1}^{L}\left|\mathsf{E}\left(\hat{I}_{k_l,i,s}-I_{k_l,i,s}\right)^{m_{k_l}}\mathsf{E}\left(\hat{I}_{k_l,j,t}-I_{k_l,j,t}\right)^{m_{k_l}}\right|\leq(2C_1\delta_{n,p})^{2L}+(n\vee p)^{-\varepsilon}.$$

it follows that

$$\sup_{(i,j)\in\mathbb{H}_0}\sup_{1\le s,t\le D}\mathsf{E}\{|\Delta_{3,ij,st}|^m\}\le C_m n^{-m/2}n^m\left\{\frac{\log(n\vee p)}{n}\right\}^m\le C_m\left[\frac{\{\log(n\vee p)\}^2}{n}\right]^{m/2}.$$

For odd integer $m$, by $\mathsf{E}\{|\Delta_{l,ij,st}|^m\}\le\mathsf{E}\{\Delta_{l,ij,st}^{m+1}\}^{m/(m+1)}$, the conclusion holds too. □

*Proof of Lemma 2.* Let the symbol "∼" stands for the asymptotic series. As an example of Laplace's method, Richter and Schumacker (1990) showed that for a random variable $X$ follows $\chi^2(K)$ for $K\in\mathbb{N}$, as $t\to\infty$,

$$\mathsf{P}(X>t)\sim\left\{\Gamma\left(\frac{K}{2}\right)\right\}^{-1}\left(\frac{t}{2}\right)^{K/2-1}e^{-t/2}\left\{1+\sum_{l=1}^{\infty}\frac{\Gamma(K/2)}{\Gamma(K/2-l)}\left(\frac{t}{2}\right)^{-l}\right\},$$

where for $K$ is a even positive integer, we put

$$\frac{\Gamma(K/2)}{\Gamma(K/2-l)}=0,\quad\text{for}\quad l>K/2.$$

Thus (A.5) holds. □

*Proof of Lemma 3.* Recall the definitions of $J_{k,ij,d}$ in (S17). And also define $J_{k,ij}$, $J_{ij}$, $V_{ij}$ and $\tilde{T}_{ij}$ the same way as in the proof of Lemma 1. Define $R_{ij}$ the same way as in (A.9).

By Chebyshev inequality and (S26) with $m=12r$,

$$\mathsf{P}(|R_{ij}|>\{\log(n\vee p)\}^{-1})\le\frac{CD^{24r}\{\log(n\vee p)\}^2n^{-6r}}{\{\log(n\vee p)\}^{-12r}}\le CD^{24r}\{\log(n\vee p)\}^{12r+2}q^{-3}.$$

It follows that

$$\begin{aligned}\mathsf{P}\{\tilde{T}_{ab}>t+\{\log(n\vee p)\}^{-1},\tilde{T}_{cd}>t+\{\log(n\vee p)\}^{-1}\}&\\ -CD^{24r}\{\log(n\vee p)\}^{12r+2}q^{-3}\le\mathsf{P}\{T_{ab}>t,T_{cd}>t\}&\\ \le\mathsf{P}\{\tilde{T}_{ab}>t-\{\log(n\vee p)\}^{-1},\tilde{T}_{cd}>t-\{\log(n\vee p)\}^{-1}\}+CD^{24r}\{\log(n\vee p)\}^{12r+2}q^{-3}.&\end{aligned}$$

When $t\le b_p+1$, $CD^{24r}\{\log(n\vee p)\}^{12r+2}q^{-3}=o[\{G_D(t)\}^2]$. Therefore

$$\begin{aligned}\mathsf{P}\{\tilde{T}_{ab}>t+\{\log(n\vee p)\}^{-1},\tilde{T}_{cd}>t+\{\log(n\vee p)\}^{-1}\}-o[\{G_D(t)\}^2]\le\mathsf{P}\{T_{ab}>t,T_{cd}>t\}&\\ \le\mathsf{P}\{\tilde{T}_{ab}>t-\{\log(n\vee p)\}^{-1},\tilde{T}_{cd}>t-\{\log(n\vee p)\}^{-1}\}+o[\{G_D(t)\}^2].\quad\text{(S14)}&\end{aligned}$$

Let $J_k = (J_{k,ab}, J_{k,cd})' \in \mathbb{R}^{2(D-1)^2}$. The expectation $\mathsf{E}(J_k) = 0$. We now derive $V = \mathsf{Var}(J_k)$.

$$\begin{aligned}
&\mathsf{Cov}\left(n^{-1/2}\sum_{k=1} J_{k,ab,g}, n^{-1/2}\sum_{k=1} J_{k,cd,h}\right) \\
&= n^{-1}\cdot \mathsf{E}\left(\sum_{k=1} J_{k,ab,g}\sum_{l=1} J_{l,ab,h}\right) \\
&= n^{-1}\sum_{k=1}^{n} \frac{\mathsf{E}\{(I_{k,a,s_g}-\nu_{s_g})(I_{k,b,t_g}-\nu_{t_g})(I_{k,c,s_h}-\nu_{s_h})(I_{k,d,t_h}-\nu_{t_h})\}}{\{\nu_{s_g}(1-\nu_{s_g})\nu_{t_g}(1-\nu_{t_g})\nu_{s_h}(1-\nu_{s_h})\nu_{t_h}(1-\nu_{t_h})\}^{1/2}}
\end{aligned}$$

When $(a,b,c,d) \in \mathbb{H}_{01}$, $\mathsf{Cov}\left(n^{-1/2}\sum_{k=1} J_{k,ab,g}, n^{-1/2}\sum_{k=1} J_{k,cd,h}\right) = 0$. Let $Z_k = (Z_{k,ab}, Z_{k,cd})'$, where $Z_{k,ab} \sim N(0, V_{ab})$ *i.i.d*, $Z_{k,cd} \sim N(0, V_{cd})$ *i.i.d*, and $Z_{k,ab}$ and $Z_{k,cd}$ independent. Let

$$\mathcal{E}_t = \{J_k, k=1,\ldots,n : (n^{-1/2}\sum_{k=1}^{n} J_{k,ij})' V_{ij}^{-1}(n^{-1/2}\sum_{k=1}^{n} J_{k,ij}) > t,\ (i,j) \in \{(a,b),(c,d)\}\}.$$

Following the similar argument as in the proof of Lemma 4, we have

$$|\mathsf{P}\{(J_1,\ldots,J_n) \in \mathcal{E}_t\} - \mathsf{P}\{(Z_1,\ldots,Z_n) \in \mathcal{E}_t\}| \le CD^6 n^{-1/2}(1+t)^{3/2}\mathsf{P}\{(Z_1,\ldots,Z_n) \in \mathcal{E}_t\}. \tag{S15}$$

Because $(n^{-1/2}\sum_{k=1}^{n} J_{k,ij})' V_{ij}^{-1}(n^{-1/2}\sum_{k=1}^{n} J_{k,ij})$, $(i,j) \in \{(a,b),(c,d)\}$ follows $\chi^2\{(D-1)^2\}$ distribution,

$$\begin{aligned}
&\left|\mathsf{P}\{\tilde{T}_{ab} > t + \{\log(n\vee p)\}^{-1}, \tilde{T}_{cd} > t + \{\log(n\vee p)\}^{-1}\} - [G_D\{t + \{\log(n\vee p)\}^{-1}\}]^2\right| \\
&\qquad \le CD^6 n^{-1/2}\{1 + t + \{\log(n\vee p)\}^{-1}\}^{3/2}[G_D\{t + \{\log(n\vee p)\}^{-1}\}]^2
\end{aligned} \tag{S16}$$

For $1 \le t < b_p$, by continuity of $G_D$,

$$G_D[t + \{\log(n\vee p)\}^{-1}]/G_D(t) \to 1.$$

Therefore, by (S14), (S16) and Lemma 1,

$$\begin{aligned}
\sup_{0\le t\le b_p+1}\ \sup_{(a,b,c,d)\in\mathbb{H}_{01}} &\Big|\mathsf{P}\{\tilde{T}_{ab} > t + \{\log(n\vee p)\}^{-1}, \tilde{T}_{cd} > t + \{\log(n\vee p)\}^{-1}\} \\
&- \mathsf{P}(T_{ab} > t)\mathsf{P}(T_{cd} > t)\Big| = o[\{G_D(t)\}^2].
\end{aligned}$$

Similarly, we can show

$$\begin{aligned}
\sup_{0\le t\le b_p+1}\ \sup_{(a,b,c,d)\in\mathbb{H}_{01}} &\Big|\mathsf{P}\{\tilde{T}_{ab} > t - \{\log(n\vee p)\}^{-1}, \tilde{T}_{cd} > t - \{\log(n\vee p)\}^{-1}\} \\
&- \mathsf{P}(T_{ab} > t)\mathsf{P}(T_{cd} > t)\Big| = o[\{G_D(t)\}^2].
\end{aligned}$$

□

*Proof of Lemma 4.* Define $J_{k,ij} = (J_{k,ij,1}, \dots, J_{k,ij,(D-1)^2})'$, where

$$J_{k,ij,d} = \frac{(I_{k,i,s_d} - \nu_{s_d})(I_{k,j,t_d} - \nu_{t_d})}{\{\nu_{s_d}(1-\nu_{s_d})\nu_{t_d}(1-\nu_{t_d})\}^{1/2}}, \quad s_d = \lceil d/(D-1) \rceil, \quad t_d = d - (s_d - 1)(D-1) + 1. \tag{S17}$$

Let $J_{ij} = n^{-1/2}\sum_{k=1}^n J_{k,ij}$, and $V_{ij} = \mathsf{Cov}(J_{k,ij})$. The proof of Theorem 1 indicates that $V_{ij}$ is a positive definite matrix with $\sup_{ij} \lambda_{\max}(V_{ij}) \le 1/C$ and $\inf_{ij} \lambda_{\min} V_{ij} > C$. Then $\tilde{T}_{ij} = (U_{ij}V_{ij}^{-1/2}n^{-1/2}\sum_{k=1}^n J_{k,ij})'(U_{ij}V_{ij}^{-1/2}n^{-1/2}\sum_{k=1}^n J_{k,ij})$, where $U_{ij}$ is an orthogonal matrix. Thus, $\tilde{T}_{ij} = (n^{-1/2}\sum_{k=1}^n J_{k,ij})'V_{ij}^{-1}(n^{-1/2}\sum_{k=1}^n J_{k,ij})$. Also $\mathsf{P}(\tilde{T}_{ij} \le t) = \mathsf{P}(J_{ij} \in \mathcal{E}_{ij,t})$, where $\mathcal{E}_{ij,t}$ is an $(D-1)^2$ dimensional ellipsoid. When $t \le C_0 \log p$, the maximum length of semi-principal axes is bounded by $C(\log p)^{1/2}$. Let $\mathcal{B}_t \subseteq \mathbb{R}^{(D-1)^2}$ be a centered ball with radius equal to $t$.

To simplify the notation, we omit $ij$ in the subscript now. Note that all the bounds shown below are uniform for all $(i,j)$, $i \neq j$.

When $J_k$ follows a non-lattice distribution, by Theorem 19.2 in Bhattacharya and Rao (2010), a bounded continous density $q_n$ of $n^{-1/2}\sum_{k=1}^n J_{k,ij}$ exists, and for all $u \in \mathbb{R}^{(D-1)^2}$,

$$|q_n(u) - \phi_V(u)| \le Cn^{-1/2}P_{1,J}(u)\phi_V(u), \tag{S18}$$

where $\phi_V(u)$ is multivariate Gaussian density function with mean 0 and covariance $V$, and $P_{1,J}(u)$ is the first Cramér-Edgeworth polynomial. The general expressions and discussions of Cramér-Edgeworth polynomials can be found in Bhattacharya and Rao (2010, Chapter 2 Section 7). It is easy to see that

$$\int_{\mathcal{E}_t^c} \phi_V(u)\,\mathrm{d}u = \int_{\mathcal{B}_t^c} \phi_I(u)\,\mathrm{d}u = G_D(t), \tag{S19}$$

Because $P_{1,J}(u)$ is a third order polynomial, by Lemma 5,

$$\left|\int_{\mathcal{E}_t^c} P_{1,J}(u)\phi_V(u)\,\mathrm{d}u\right| = \left|\int_{\mathcal{B}_t^c} P_{1,V^{-1/2}J}(u)\phi_I(u)\,\mathrm{d}u\right| \le C(1+t)^{3/2}(D-1)^6\{G_D(t)\} \tag{S20}$$

Combine (S18) – (S20), we have (A.7) holds.

Next, consider the case when $J_k$ follows a lattice distribution. Let $\mathbb{L} = \{\alpha \in \mathbb{R}^{(D-1)^2} : \mathsf{P}(\sum_{k=1}^n J_k = u) > \alpha\}$. Also, let $u_{\alpha,n} = n^{-1/2}\alpha$, and $p_n(u_{\alpha,n}) = \mathsf{P}(\sum_{k=1}^n J_k = \alpha)$. Denote by $l = (l_1, \ldots, l_{(D-1)^2})$ the span (the distance between two closest points in each direction) of $\mathbb{L}$. Let $L = \prod_{d=1}^{(D-1)^2} l_d$.

Define $\psi_{V,0}(u) = Ln^{-(D-1)^2/2}\phi_V(u)$, and $\psi_{V,J,1}(u) = Ln^{-(D-1)^2/2}P_{1,J}(u)\phi_V(u)$. For positive integer $m$, let $\mathcal{A}_{t,m} = \mathcal{B}_t^c \cap \{u_{\alpha,n} : V^{1/2}\alpha \in \mathbb{L}\}$. And let $\mathcal{F}_{t,m} = \{u : V^{-1/2}u \in \mathcal{A}_{t,m}\}$. It is easy to see that $\mathcal{F}_{t,m} = \mathcal{E}_t^c \cap \{u_{\alpha,n} : \alpha \in \mathbb{L}\}$.

By Theorem 22.1 in Bhattacharya and Rao (2010),

$$\sum_{u \in \mathcal{F}_{t,m}} |p_n(u) - \psi_{V,0}(u)| \leq Cn^{-1/2} \sum_{u \in \mathcal{F}_{t,m}} \psi_{V,J,1}(u).$$

Therefore,

$$\left| \sum_{u \in \mathcal{F}_{t,m}} p_n(u_{\alpha,n}) - \sum_{u \in \mathcal{A}_{t,m}} \psi_{I,0}(u) \right| \leq Cn^{-1/2} \sum_{u \in \mathcal{A}_{t,m}} \psi_{I,V^{-1/2}J,1}(u).$$

By the Euler-Maclaurin summation formula (Theorem A.4.2 in Bhattacharya and Rao (2010)),

$$\left| \sum_{u \in \mathcal{F}_{t,m}} p_n(u_{\alpha,n}) - \int_{\mathcal{B}_t^c} \phi_I(u)\,\mathrm{d}u \right| \leq Cn^{-1/2} \int_{\mathcal{B}_t^c} P_{1,V^{-1/2}J}(u)\phi_I(u)\,\mathrm{d}u.$$

Following the similar argument for the non-lattice distribution case, we have (A.7) holds for the lattice distribution case too. □

*Proof of Lemma 5.* Define the set $\mathbb{K}_D = \{k = (k_1, \ldots, k_D)' : \sum_{d=1}^D k_d \leq K\}$. Write

$$Q_K(u) = \sum_{k \in \mathbb{K}_D} c_k \prod_{d=1}^D (-1)^{k_l} H_{k_l}(u_d),$$

where $H_{k_l}$ is the $k_l$-th order Hermite polynomial, and $c_k \leq C$. Note that $\phi_1^{(k_l)}(u_d) = (-1)^{k_l} H_{k_l}(u_d)\phi_1(u_d)$.

We will prove that for any $t \geq 0$ and $k = (k_1, \ldots, K_D) \in \mathbb{K}_D$,

$$\left| \int_{\|u\|^2 \geq t} \prod_{d=1}^D \phi^{(k_l)}(u_d)\,\mathrm{d}u \right| \leq C_K (1+t)^{K/2} \int_t^\infty f_{\chi_D^2}(x)\,\mathrm{d}x. \tag{S21}$$

Let's first assume (S21) holds. Then,

$$\left|\int_{\|u\|^2\geq t} Q_K(u)\phi_{I_D}(u)\,\mathrm{d}u\right| = \sum_{k\in\mathbb{K}_D}\left|c_k\int_{\|u\|^2\geq t}\prod_{d=1}^{D}\phi^{(k_l)}(u_d)\,\mathrm{d}u\right| \leq C_K D^K(1+t)^{K/2}\int_t^\infty f_{\chi^2_D}(x)\,\mathrm{d}x.$$

Thus, Lemma 5 holds.

Let's get back to prove (S21). When $D=1$,

$$\left|\int_{u^2\geq t}\phi_1^{(k_1)}(u)\,\mathrm{d}u\right| \leq 2\Phi_1^{(k_1)}(-\sqrt{t}) \leq C_K(1+t)^{K/2}\Phi_1(-\sqrt{t}) = C_K(1+t)^{K/2}\int_t^\infty f_{\chi^2_1}(x)\,\mathrm{d}x.$$

Now suppose (S21) holds for $1,\ldots,D-1$. Let $\tilde{u}=(u_2,\ldots,u_D)'$.

$$\int_{\|u\|^2\geq t}\prod_{d=1}^{D}\phi^{(k_l)}(u_d)\,\mathrm{d}u =$$

$$\int_{\|\tilde{u}\|^2\geq t}\prod_{d=2}^{D}\phi^{(k_l)}(u_d)\,\mathrm{d}\tilde{u} + \int_{\|\tilde{u}\|^2\leq t}2\Phi_1^{(k_1)}(-\sqrt{t-\|\tilde{u}\|^2})\prod_{d=2}^{D}\phi^{(k_l)}(u_d)\,\mathrm{d}\tilde{u} \quad \text{(S22)}$$

Because $\tilde{u}\in\mathbb{R}^{(D-1)}$,

$$\left|\int_{\|\tilde{u}\|^2\geq t}\prod_{d=2}^{D}\phi^{(k_l)}(u_d)\,\mathrm{d}\tilde{u}\right| \leq \left|\int_{\|\tilde{u}\|^2\geq t}\prod_{d=2}^{D}u_d^{k_d}\phi_1(u_d)\,\mathrm{d}\tilde{u}\right| \leq C_K(1+t)^{K/2}\int_t^\infty f_{\chi^2_D}(x)\,\mathrm{d}x \quad \text{(S23)}$$

The other term

$$\begin{aligned}&\int_{\|\tilde{u}\|^2\leq t}2\Phi_1^{(k_1)}(-\sqrt{t-\|\tilde{u}\|^2})\prod_{d=2}^{D}\phi^{(k_l)}(u_d)\,\mathrm{d}\tilde{u}\\ \leq& C_K\int_{\|\tilde{u}\|^2\leq t}(t-\|\tilde{u}\|^2)^{k_1/2}\prod_{d=2}^{D}u_d^{k_d}2\Phi_1(-\sqrt{t-\|\tilde{u}\|^2})\phi_1(u_d)\,\mathrm{d}\tilde{u}\\ \leq& C_K(1+t)^{K/2}\int_{\|\tilde{u}\|^2\leq t}\Phi_1(-\sqrt{t-\|\tilde{u}\|^2})\prod_{d=2}^{D}\phi_1(u_d)\,\mathrm{d}\tilde{u}.\end{aligned}$$

On the other hand,

$$\int_t^\infty f_{\chi^2(D)}(x)\,\mathrm{d}x = \int_{\|\tilde{u}\|^2\geq t}\prod_{d=2}^{D}\phi_1(u_d)\,\mathrm{d}\tilde{u} + \int_{\|\tilde{u}\|^2\leq t}2\Phi_1(-\sqrt{t-\|\tilde{u}\|^2})\prod_{d=2}^{D}\phi_1(u_d)\,\mathrm{d}\tilde{u}.$$

Therefore,

$$\int_{\|\tilde{u}\|^2\leq t}2\Phi_1^{(k_1)}(-\sqrt{t-\|\tilde{u}\|^2})\prod_{d=2}^{D}\phi^{(k_l)}(u_d)\,\mathrm{d}\tilde{u} \leq C_K(1+t)^{K/2}\int_t^\infty f_{\chi^2_D}(x)\,\mathrm{d}x. \quad \text{(S24)}$$

Combining (S22)-(S24), we have (S21). □

*Proof of Lemma 6.* As in Lemma 1 let $\mathcal{M}_L = \{(m_1, \ldots, m_n) : \sum_{k=1}^{n} I(m_k \ge 1) = L\}$. For any $(m_1, \ldots, m_n) \in \mathcal{M}_L$, let $(k_1, \ldots, k_L) = \{k : m_k \ge 1\}$. Under $\mathrm{H}_{0,ij}$, $\mathsf{E}(L_{ij,st}) = 0$, where $L_{ij,st}$ is defined in (A.8). When $m$ is even,

$$\sup_{1 \le s,t \le D} \mathsf{E}(L_{ij,st}^m) = n^{-m/2}(\nu_s \nu_t)^{-m/2} \times \sum_{1 \le L \le m} \sum_{m_1,\ldots,m_n \in \mathcal{M}_L} \frac{m!}{m_1! \ldots m_n!} \prod_{l=1}^{L} \mathsf{E}\left\{(I_{k_l,i,s} - \nu_s)^{m_{k_l}}\right\} \mathsf{E}\left\{(I_{k_l,j,t} - \nu_t)^{m_{k_l}}\right\}. \quad \text{(S25)}$$

By (S11) in the supplementary material,

$$\sup_{1 \le s,t \in D} \mathsf{E}(L_{ij,st}^m) \le C_m n^{-m/2} n^{m/2} = C_m,$$

where $C_m = \{4(1+\varepsilon)C_1C_2\}^m \mu_0^{-m} m!$. By Lemma 1, $\sup_{1 \le s,t \le D} \mathsf{E}(\Delta_{ij,st}^2) \le C_m[\{\log(n \vee p)\}^2/n]^{m/2}$. Then

$$\mathsf{E}|R_{ij}|^m \le C_m D^{2m} \sup_{1 \le s,t \le D} (\mathsf{E}L_{ij}^{2m})^{1/2} (\mathsf{E}\Delta_{ij,st}^{2m})^{1/2} + C_m D^{2m} \sup_{1 \le s,t \le D} \mathsf{E}(\Delta_{ij,st}^{2m}) \le C_m D^{2m} \{\log(n \vee p)\}^m n^{-m/2}. \quad \text{(S26)}$$

By (S26) and the Markov Inequality, for any constant $c > 0$

$$\mathsf{P}(|R_{ij}| > \{\log(n \vee p)\}^{-1}) \le C_m D^{2m} \{\log(n \vee p)\}^{2m} n^{-m/2}.$$

When $0 < t \le 2M^2$, for some positive constant $M$, we can take sufficiently large $m$ so that as $n \to \infty$, $\mathsf{P}(|R_{ij}| > \{\log(n \vee p)\}^{-1})/G_D(t) \le \varepsilon_p$.

As $2M^2 < t \le C_0 \log(n \vee p)$, by Lemma 2, we have

$$\begin{aligned}
\frac{\mathsf{P}(|R_{ij}| > \{\log(n \vee p)\}^{-1})}{G_D(t)} &\le \frac{C_m D^{2m} \{\log(n \vee p)\}^{2m} n^{-m/2}}{\Gamma\left(\frac{(D-1)^2}{2}\right)^{-1} M^{(D-1)^2-2} \exp\{-C_0 \log(n \vee p)/2\}} \\
&\le C_m M^{-(D-1)^2+2} D^{2m} \Gamma\left(\frac{(D-1)^2}{2}\right) \{\log(n \vee p)\}^{2m} n^{-2m+C_0 r/2} \\
&\le C_m M^{-(D-1)^2+2} D^{2m} \left\lfloor \frac{(D-1)^2}{2} \right\rfloor ! \{\log(n \vee p)\}^{2m} n^{-2m+C_0 r/2}.
\end{aligned}$$

By Sterling inequality and the condition, $D \le C_3 \left\{ \frac{\log(n\vee p)}{\log\log(n\vee p)} \right\}^{1/2}$, Then

$$
\begin{aligned}
&\frac{\mathsf{P}(|R_{ij}| > \{\log(n \vee p)\}^{-1})}{G_D(t)} \\
&\le C C_m M^{-(D-1)^2+2} \cdot \left\{ \frac{\log(n \vee p)}{\log\log(n \vee p)} \right\}^{m+1/2} \\
&\quad \cdot \exp\left\{ C \frac{\log(n \vee p)}{\log\log(n \vee p)} + \log(n \vee p) - \frac{\log(n \vee p) \log\log\log(n \vee p)}{\log\log(n \vee p)} \right\} \\
&\quad \cdot \{\log(n \vee p)\}^{2m} n^{-2m+C_0 r/2} \\
&\le C C_m M^{-(D-1)^2+2} \left\{ \frac{\log(n \vee p)}{\log\log(n \vee p)} \right\}^{m+1/2} \{\log(n \vee p)\}^{2m} n^{-2m+(C_0+2)r/2} \qquad \text{(S27)}
\end{aligned}
$$

For any constant $C_0$, we can take sufficiently large $m$ such that $-2m + (C_0+2)r/2 < 0$. Then the leading term of (S27) is $n^{-2m+(C_0+2)r/2}$. Then for $2M^2 < t \le C_0 \log(n \vee p)$, $n \to \infty$, $\mathsf{P}(|R_{ij}| > \{\log(n \vee p)\}^{-1})/G_D(t) \le \varepsilon_p$. □

*Proof of Lemma 7.* We consider the Graph $G_{abcd} = (V_{abcd}, E_{abcd})$, where $V_{abcd}$ is the vertex set and $E_{abcd}$ is the edge set. When $(a,b) \ne (c,d)$, $V_{abcd}$ could contain 3 or 4 vertices. There is an edge between $i \ne j \in V_{abcd}$ if and only if $Y_i \not\perp\!\!\!\perp Y_j$. If $\mathsf{Card}(V_{abcd}) = v$ and $\mathsf{Card}(E_{abcd}) = e$, we call $G_{abcd}$ a $vVeE$ graph. We say $G_{abcd}$ satisfies $(\star)$ if

$$G_{abcd} \text{ is a } 4V \text{ graph and there is at least one isolated vertex; or } G_{abcd} \text{ is } 3V0E. \qquad (\star)$$

Divide the set $\mathbb{H}_{00} = \{(a,b,c,d) : (a,b) \in \mathbb{H}_0, (c,d) \in \mathbb{H}_0, (a,b) \ne (c,d)\}$ into two sets:

$$\mathbb{H}_{01} = \{(a,b,c,d) \in \mathbb{H}_0 : (a,b) \ne (c,d) \in \mathbb{H}_0, G_{abcd} \text{ satisfies } (\star)\}, \quad \mathbb{H}_{02} = \mathbb{H}_{00} \setminus \mathbb{H}_{01}. \qquad \text{(S28)}$$

On $\mathbb{H}_{02}$, $G_{abcd}$ is $3V1E$, $4V2E$, $4V3E$ or $4V4E$.

It suffices to prove that for $\mathbb{H}_{0l}$, $l = 1, 2$,

$$\sup_{0 \le t \le b_p} \frac{1}{q^2 \{G_D(t)\}^2} \sum_{(a,b,c,d) \in \mathbb{H}_{0l}} \{\mathsf{P}(T_{ab} > t, T_{cd} > t) - \mathsf{P}(T_{ab} > t)\mathsf{P}(T_{cd} > t)\} \to 0. \qquad \text{(S29)}$$

We first show (S29) holds for $\mathbb{H}_{02}$. Because

$$|\mathsf{P}(T_{ab} > t, T_{cd} > t) - \mathsf{P}(T_{ab} > t)\mathsf{P}(T_{cd} > t)| \le 2G_D(t), \quad \mathsf{Card}(\mathbb{H}_{02}) \le p d_p^2 + p^2 d_p^2 + p d_p^3,$$

$$\frac{1}{q^2\{G_D(t)\}^2}\sum_{(a,b,c,d)\in\mathbb{H}_{02}}\{\mathsf{P}(T_{ab}>t,T_{cd}>t)-\mathsf{P}(T_{ab}>t)\mathsf{P}(T_{cd}>t)\}$$

$$\leq C\frac{p^2d_p^2G_D(t)}{q^2\{G_D(t)\}^2}=C\frac{d_p^2}{qG_D(t)}. \quad \text{(S30)}$$

When $0\leq t\leq b_p$, $qG_D(t)\geq \alpha c_p$, then ($S30$) $\leq d_p^2/c_p\to 0$.

By Lemma 3 and $\mathsf{Card}(\mathbb{H}_{01})\leq q_0^2$, we have (S29) holds for $\mathbb{H}_{01}$. □

## S3. Detailed description of the six settings in the numerical experiments

The dependence between $Y_i$ and $Y_j$ is determined by the dependence between $U_{0,i}$ and $U_{0,j}$. We consider the following 5 settings.

**SE1** Linear dependence, Gaussian-tail. We first generate $(Y_{0,1},\ldots,Y_{0,p})'$ from multivariate Gaussian distribution $\mathrm{N}(0,\mathbf{\Sigma})$, and let $U_{0,i}=\Phi(Y_{0,i})$. We set $\mathbf{\Sigma}$ as a block matrix $\mathsf{diag}(\mathbf{\Sigma}_1,\mathbf{\Sigma}_2,\mathbf{\Sigma}_3)$, where each $\mathbf{\Sigma}_s$ has column numbers and row numbers equal to $p_s$, $s=1,2,3$. Let $p_1=5$ and $p_2=40$. Then $p_3=p-p_1-p_2$. Construct $\mathbf{\Sigma}_1$ by first generating a random matrix $\mathbf{M}_1$ with diagonal 0 and non-diagonal elements following Unif$(0.5,0.6)$, and then let $\tilde{\mathbf{M}}_1=\mathbf{M}_1+t(\mathbf{M}_1)$. Next, we set $\tilde{\mathbf{\Sigma}}_1=\tilde{\mathbf{M}}_1+a\mathbf{I}_{p_1\times p_1}$ for some constant $a$ so that $\tilde{\mathbf{M}}_1$ is a positive definite matrix with the condition number (the ratio of largest eigenvalue and the smallest eigenvalue) equal to 100. Then $\tilde{\mathbf{\Sigma}}_1$ is standardized so that the diagonal elements are all 1. Denote the resulting matrix by $\mathbf{\Sigma}_1$. This means that $(Y_{0,1},\ldots,Y_{0,p_1})'$ are mutually dependent. The second block $\mathbf{\Sigma}_2$ is a block diagonal matrix of $p_2/2$ block matrices with dimension $2\times 2$. This means that for $i=p_1+1,p_1+3,\ldots,p_1+p_2-1$, $Y_{0,i}$ and $Y_{0,i+1}$ are dependent. We generate the correlation between $Y_{0,i}$ and $Y_{0,i+1}$ from $0.5\mathrm{Unif}(0.2,0.6)+0.5\mathrm{Unif}(-0.6,-0.2)$. The third block $\mathbf{\Sigma}_3$ is identity matrix, so that for $i>p_1+p_2$, $Y_{0,i}$ is independent of all other variables. Overall, among total $p(p-1)/2$ pairs, thirty of them are conditionally dependent, and the rest are conditionally independent.

**SE2** Linear dependence with outliers, Gaussian-tail. We first adopt the same dependence structure for $(Y_{0,1},\ldots,Y_{0,p})'$ as that in Setting 1. Then we contaminate the sam-ples $(Y_{0,1},\ldots,Y_{0,p})'$ by 10% weak outliers, generated independently from the centered Cauchy distribution.

**SE3 & SE6** Quadratic dependence. We set the first 30 pairs $(Y_1,Y_2)$, $(Y_3,Y_4)$, ..., $(Y_{59},Y_{60})$ are dependent, and all other pairs are independent. For $i>60$, we generate $U_{0,i}$ independently from Unif$(0,1)$ distribution. For $i=1,3,\ldots,59$, we generate $U_{0,i}$ from Unif$(0,1)$ distribution, and set $Z_i=\Phi^{-1}(U_{0,i})$. We then Let $Z_{i+1}=Z_i^2+E_i$, where $E_i$ independently follows $\chi^2(1)$ distribution. Thus, $Z_{i+1}$ marginally follows $\chi^2(2)$ distribution. Although $Z_i$ and $Z_{i+1}$ are dependent, their linear dependence $\mathsf{Cov}(Z_i,Z_{i+1})=0$. Let $U_{0,i+1}=F_{\chi^2(2)}(Z_{i+1})$, then $U_{0,i+1}$ follows Unif$(0,1)$ distribution. Quadratic dependence is an example of non-linear dependence. We then set $Y_{0,i}=F_Y(U_{0,i})$. In **SE3**, $F_Y$ is the CDF of standard Gaussian; and in **SE6**, $F_Y$ is the CDF of standard Cauchy.

**SE4** Dependence affected by latent variables, Gaussian-tail. In many applications, the dependence between two variables depends on other latent variables, which are unfortunately not observed and not included in the model. It is important to check if a method can identify dependent paris for this case. We set $p_1=5$, $p_2=40$, and $p_3=p-p_1-p_2$, so that the first $p_1$ variables $(U_{0,1},\ldots,U_{0,p_1})$ are mutually dependent, the next $p_2/2$ pairs $(U_{0,p_1+1},U_{0,p_1+2})$, $(U_{0,p_1+3},U_{0,p_1+4})$, ..., $(U_{0,p_1+p_2-1},U_{0,p_1+p_2})$ are dependent, and all other pairs are independent. We first generate $Z_0$ from standard Gaussian distribution. For $i=1,\ldots,p_1$, generate a latent variable $L_i$ from Unif$(-1,1)$ distribution and an error term $E_i$ from standard Gaussian. Then we set $Z_i=(16L_i^2+1)^{-1/2}(4L_iZ_0+E_i)$. Then $Z_i$ follows the standard Gaussian distribution. Conditioning on the latent variable $L_i$, the dependence between $Z_i$ and $Z_0$ is linear with $\mathsf{Cov}(Z_i,Z_0\mid L_i)=(16L_i^2+1)^{-1/2}4L_i$. For $i=p_1+1,p_1+3,\ldots,p_1+p_2-1$, generate $Z_i$ and $E_i$ from standard Gaussian, and set $Z_{i+1}=(16L_i^2+1)^{-1/2}(4L_iZ_i+E_i)$. For $i=1,\ldots,p_1+p_2$, set $U_{0,i}=\Phi^{-1}(Z_i)$ to generate uniform random variables. For $i>p_1+p_2$, generate $U_{0,i}$ independently from Unif$(0,1)$ distribution.

**SE5** Dependence affected by marginal values of variables, Gaussian-tail. In genomic studies,

the gene association could be affected by the expression level of the variables. For example, Gene A and its protein product may promote the expression of Gene B when its expression level is high, do not affect Gene B when its expression level is moderate, and depress the expression of Gene B when its expression level is low. In this case, the dependence of Gene A and Gene B is affected by the marginal expression level of Gene A. Similar as Setting 4, we set $p_1 = 5$, $p_2 = 40$, and $p_3 = p - p_1 - p_2$, so that the first $p_1$ variables are mutually dependent, the next $p_2/2$ pairs are dependent within pairs, and all other pairs are independent. We first generate $Z_0$ from standard Gaussian distribution. And for $i = 1, \ldots, p_1$, we generate $E_i$ from standard Gaussian too. Set

$$Y_{0,i} = \begin{cases} \frac{\sqrt{63}}{8} Z_0 + \frac{1}{8} E_i & \text{if } Z_0 > 1; \\ E_i & \text{if } -1 \leq Z_0 \leq 1; \\ -\frac{\sqrt{63}}{8} Z_0 + \frac{1}{8} E_i & \text{otherwise.} \end{cases}$$

For $i = p_1 + 1, p_1 + 3, \ldots, p_1 + p_2 - 1$, we generate $Y_{0,i}$ and $E_i$ independently from standard Gaussian distribution, and let

$$Y_{0,i+1} = \begin{cases} \frac{\sqrt{63}}{8} Y_{0,i} + \frac{1}{8} E_i & \text{if } Y_i > 1; \\ E_i & \text{if } -1 \leq Y_i \leq 1; \\ -\frac{\sqrt{63}}{8} Y_{0,i} + \frac{1}{8} E_i & \text{otherwise.} \end{cases}$$

For $i = 1, \ldots, p_1 + p_2$, $Y_{0,i}$ marginally follows standard Gaussian distribution. Therefore, $U_{0,i} = \Phi(Y_{0,i})$ follows $\text{Unif}(0, 1)$ distribution. For $i > p_1 + p_2$, generate $U_{0,i}$ independently from $\text{Unif}(0, 1)$ distribution.

## S4. Detailed description of the multiple testing methods kendall and spearman

The Kendall's $\tau$ coefficient is $\tau_{ij} = 2(N_c - N_d)/\{n(n-1)\}$, where

$$N_c = \sum_{k=1}^{n} \sum_{l \neq k} I(Y_{0,k,i} < Y_{0,l,i}, Y_{0,k,j} < Y_{0,l,j}), \quad N_d = n(n-1)/2 - N_c$$

are the number of concordant and discordant pairs. Let the standardized test statistic

$$\hat{\tau}_{ij} = \tau_{ij} \left\{ \frac{9n(n-1)}{2(2n+5)} \right\}^{1/2}$$

Under $\mathrm{H}_{0,ij}$, the statistic $\hat{\tau}_{ij}$ asymptotically follows the standard Gaussian distribution. Similar to our proposed multiple testing procedure, we develop the following procedure

$$\text{Reject } \mathrm{H}_{0,ij} \text{ if } |\hat{\tau}_{ij}| \geq \hat{t}, \quad \hat{t} = \inf \left\{ 0 \leq t \leq b_p : \frac{G_D(t)(p^2-p)/2}{\max\left\{\sum_{1\leq i<j\leq p} I(|\hat{\tau}_{ij}| \geq t), 1\right\}} \leq \alpha \right\}. \quad \text{(S1)}$$

Here $G_D(t) = 2\{1 - \Phi(-t)\}$, and $b_p = (4 \log p - \log_2 p - \log_3 p)^{1/2}$. We denote this method by KENDALL.

The Spearman's $\rho$ coefficient is

$$\rho_{ij} = 1 - \frac{6 \sum_{k=1}^{n} d_k}{n(n^2-1)}, \quad d_k = \text{RANK}(Y_{0,k,i}) - \text{RANK}(Y_{0,k,j}).$$

Here RANK is the tied rank operator. The transformed statistic $\hat{\rho}_{ij} = (n-2)^{1/2} \rho_{ij} (1-\rho_{ij}^2)^{-1/2}$ asymptotically follows $T(n-2)$ distribution. Therefore, we use the testing procedure (S1) with $G_D(t) = 2\{1 - F_{T(n-2)}(t)\}$ distribution and $\hat{\tau}_{ij}$ replaced by $\hat{\rho}_{ij}$. We denote this method by SPEARMAN.

## S5. Why the Adaptive Lasso Method Developed in Fan et al. (2014) Yields Estimates that Satisfy Condition C2*?

Fan et al. (2014) proposed an adaptive robust Lasso method such that under certain conditions (discussed in Fan et al. (2014)), with probability at least $1 - n^{-(2+\varepsilon)r}$,

$$\|\hat{\boldsymbol{\beta}}_{1,i}(\tau) - \boldsymbol{\beta}_{1,i}(\tau)\|_2 \leq C_4 \{s_x (\log n)/n\}^{1/2}, \quad \hat{\boldsymbol{\beta}}_{2,i}(\tau) = \mathbf{0}. \quad \text{(S1)}$$

Here $\boldsymbol{\beta}_{1,i}(\tau)$ is the vector of non-zero coordinates of $\boldsymbol{\beta}_{0,i}(\tau)$, and $\hat{\boldsymbol{\beta}}_{1,i}$ is the corresponding estimates. $\hat{\boldsymbol{\beta}}_{2,i}$ is the estimator of the zero coordinates of $\boldsymbol{\beta}_{0,i}(\tau)$.

We can also partition the covariates $\boldsymbol{X}$ into two parts, $\boldsymbol{S}$ and $\boldsymbol{Q}$, where $\boldsymbol{S}$ is the subvector of $\boldsymbol{X}$ corresponding to the covariates whose coefficients are non-zero, and $\boldsymbol{Q}$ is the remaining part. Assume $\sup_{k=1}^{n}\|\boldsymbol{S}_k\|_2 \le C_5$, by (S1), we have

$$\begin{aligned}\mathsf{P}\left[|\hat{Q}_{k,i,s} - Q_{k,i,s}| > 2C_2(1+\varepsilon)n^{-1/2}\{\log(n\vee p)\}\right] \\ \le \mathsf{P}\left\{\|\hat{\boldsymbol{\beta}}_{1,i}(\tau_s) - \boldsymbol{\beta}_{1,i}(\tau_s)\|_2 > \frac{2C_2 r}{C_5}(1+\varepsilon)n^{-1/2}\log n\right\}.\end{aligned}$$

If we take $C_2 > C_3 C_4 s_x^{1/2}/\{2(1+\varepsilon)\}$, then

$$\mathsf{P}\left[|\hat{Q}_{k,i,s} - Q_{k,i,s}| > 2C_2(1+\varepsilon)n^{-1/2}\{\log(n\vee p)\}\right] < (n\vee p)^{-2+\varepsilon}.$$

This then leads to (15).